\documentclass[aps,prd,showkeys,nofootinbib,superscriptaddress]{revtex4-2}
\usepackage{dsfont,tensor,epsf}
\usepackage{graphicx,pslatex}

\usepackage{adjustbox}

\usepackage{hyperref}
\usepackage{color}
\definecolor{darkgreen}{rgb}{0,0.35,0}

\usepackage{amsfonts}
\usepackage{amsmath}
\usepackage{subfig}
\usepackage{mathtools,slashed,tensor}

\usepackage{tikz}
\usepackage[compat=1.1.0]{tikz-feynman}
\usepackage{xcolor}
\usetikzlibrary{decorations.pathmorphing}
\usetikzlibrary{decorations.markings}
\usetikzlibrary{decorations.pathreplacing}

\usepackage{widebar}
\usepackage{physics}
\usepackage[utf8]{inputenc}
\usepackage[english]{babel}
\usepackage{csquotes}
\usepackage{amsfonts}
\usepackage{amssymb}
\usepackage{xparse}
\usepackage{ifthen}

\newcommand{\dual}[1]{\widetilde{#1}}
\newcommand{\DD}[1]{\mathcal{D}#1\,}
\DeclareDocumentCommand\diag{}{\operatorname{diag}\quantity}
\DeclareDocumentCommand\polylog{m}{\operatorname{Li}_{#1}\!\quantity}
\renewcommand{\vec}[1]{\mathbf{#1}}
\renewcommand{\bar}[1]{\widebar{#1}}
\renewcommand{\tilde}[1]{\widetilde{#1}}

\graphicspath{{../../figures/}}

\newcommand{\cecs}{Centro de Estudios Cient\'{\i}ficos (CECs), Casilla 1469, Valdivia, Chile}
\newcommand{\uss}{Universidad San SebastiÃ¡n, Facultad de Ingenier a, Arquitectura y DiseÃ±o, sede Valdivia,
General Lagos 1163, Valdivia 5110693, Chile}
\newcommand{\ucharles}{Faculty of Mathematics and Physics, Charles University, V Hole\v{s}ovi\v{c}k\'ach 2, 18000 Prague 8, Czech Republic}
\newcommand{\kulak}{Department of Physics, KU Leuven Campus Kortrijk--Kulak, Etienne Sabbelaan 53 bus 7657, 8500 Kortrijk, Belgium}
\newcommand{\infn}{ INFN, Sezione di Napoli, Complesso Universitario di Monte S.~Angelo, Via Cintia Edificio 6, 80126 Naples, Italia}
\newcommand{\napoli}{Dipartimento di Matematica e Applicazioni "R. Caccioppoli", Universit\'{a} di Napoli Federico II, Complesso Universitario di Monte S.~Angelo,  Via Cintia Edificio 6,
80126 Naples, Italia }
\newcommand{\uach}{Instituto de Ciencias F\'{i}sicas y Matem\'{a}ticas, Universidad Austral de Chile, Casilla 567, Valdivia, Chile}
\newcommand{\ughent}{Ghent University, Department of Physics and Astronomy, Krijgslaan 281-S9, 9000 Ghent, Belgium}

\tikzset{
    /tikzfeynman/warn luatex=false,
    photon/.style={
        /tikzfeynman/photon
    },
    double photon/.style={
        double,
        double distance=2pt,
        /tikzfeynman/photon
    },
    scalar/.style={
        /tikzfeynman/scalar
    }
}

\DeclareDocumentCommand{\oneloop}{m m m O{1cm} O{.5cm} O{0}}{
    \begin{tikzpicture}[baseline=0cm]
        \draw[#1] (0,0) circle[radius=#5];

        \ifthenelse{#3=0}{
        }{
            \pgfmathsetmacro\angleshift{360/#3}
            \foreach \i in {0, 1, ..., #3} {
                \pgfmathsetmacro\phi{\angleshift*\i + #6}
                \draw[#2] (\phi:#5) -- +(\phi:#4);
            }
        }
    \end{tikzpicture}
}

\DeclareDocumentCommand{\oneloopn}{m m O{1cm} O{.5cm}}{
    \begin{tikzpicture}[baseline=0cm]
        \pgfmathsetmacro\gap{45}
        \pgfmathsetmacro\aoffset{-40}
        \pgfmathsetmacro\startangle{\gap+\aoffset}
        \pgfmathsetmacro\stopangle{360+\aoffset}
        \pgfmathsetmacro\totangle{360-\gap}

        \draw[#1] (0,0) circle[radius=#3];

        \pgfmathsetmacro\angleshift{\totangle/4}
        \foreach \i in {0, 1, ..., 2} {
            \pgfmathsetmacro\phi{\angleshift*(\i+3/2) + \startangle}
            \pgfmathsetmacro\n{int(\i+1)}
            \draw[#2] (\phi:#3) -- ++(\phi:#4) ++(\phi:.5em) node[anchor=center] {\(\n\)};
        }

        \pgfmathsetmacro\midgap{\stopangle+\gap/2}
        \foreach \i in {-1, 0, 1} {
            \pgfmathsetmacro\phi{\midgap + \i*\gap/4}
            \fill[black] (\phi:#3+#4/2) circle[radius=.25mm];
        }

        \pgfmathsetmacro\phi{\angleshift/2 + \startangle}
        \draw[#2] (\phi:#3) -- ++(\phi:#4) ++(\phi:.5em) node[anchor=center] {\(n\)};
    \end{tikzpicture}
}

\begin{document}

\title{The Casimir effect in chiral media using path integral techniques}

\author{Fabrizio Canfora}
\email{fabrizio.canfora@uss.cl}
\affiliation{\uss}
\affiliation{\cecs}
\author{David Dudal}
\email{david.dudal@kuleuven.be}
\affiliation{\kulak}
\affiliation{\ughent}
\author{Thomas Oosthuyse}
\email{thomas.oosthuyse@kuleuven.be (corresponding author)}
\affiliation{\kulak}
\author {Pablo Pais}
\email{pais@ipnp.mff.cuni.cz}
\affiliation{\ucharles}
\affiliation{\uach}
\author{Luigi Rosa}
\email{rosa@na.infn.it}
\affiliation{\napoli}
\affiliation{\infn}

\begin{abstract}
    We employ path integral methods to calculate the Casimir energy and force densities in a chiral extension of QED.
    Manifestly gauge invariant perfect electromagnetic boundary conditions, a natural generalization of perfect electric and perfect magnetic conditions,
    are implemented directly in the action by the usage of auxiliary fields.
    The chiral properties of the vacuum are modelled using a background $\theta$ field,
    and we introduce techniques to efficiently calculate the path integral in this chiral medium.
    The flexibility of our method allows us to naturally obtain results for a variety of configurations,
    and where comparison is possible our results are in perfect agreement with existing literature.
    Among these are multiple situations where a repulsive Casimir force is possible.
\end{abstract}

\maketitle

\section{Motivation}

The Casimir effect, and its associated force, describes how even neutral objects experience forces through their disturbance of the vacuum structure.
This phenomenon is well known in QED, and has been verified in a multitude of experiments, see e.g.\ \cite{Plunien:1986ca,Mohideen:1998iz,Lambrecht:1999vd,Bordag:2001qi,Bressi:2002fr,Milton:2004ya,Bimonte:2021sib}.
While the Casimir effect is not necessarily proof of the physicality of vacuum energy \cite{Jaffe:2005vp},
the formulation in terms of vacuum energies remains convenient for practical calculations.

The Casimir effect is not only of theoretical interests, it also plays an important role in the production and operation of micro(electro)mechanical systems,
as on the nanometer scale the Casimir force is non-negligible and needs to be taken into account \cite{genet2008casimir,Chan:2001zza,Chan:2001zzb,serry1998role}.
In such situations it is beneficial to be able to tune the Casimir force to a desired strength to create novel applications, e.g\ with a repulsive Casimir force.
A repulsive Casimir force is however only possible in the case that reflection symmetry is broken, see the no-go theorem of \cite{Kenneth:2006vr}.
Such a broken reflection symmetry can be obtained by using a geometrically non-symmetric setup,
but it is also possible to break reflection symmetry in the vacuum (or better said, medium) itself while the geometry remains symmetric.
There is also a third possibility of applying boundary conditions (e.g.~at plates) which break reflection symmetry.
We will use a combination of non-symmetric boundary conditions and medium, namely a chiral medium, to obtain a repulsive Casimir force in this paper.

There is also growing proof from lattice simulations that the Casimir effect has nontrivial consequences in \((1+1)\)d \(\mathbb{C}P^{N-1}\) and \((2+1)\)d Yang-Mills models \cite{Chernodub:2018pmt,Chernodub:2019nct}.
Specifically in \((2+1)\)d Yang-Mills theory, there seems to be an interesting interplay between the Casimir effect, the deconfinement transition,
and a dynamically generated mass scale which differs from the lowest glueball mass. Moreover, the QCD quark-gluon plasma has also been motivated to form a chiral medium with asymmetric behaviour for left-and right quarks, potentially leading to novel phenomena \cite{Kharzeev:2013ffa}.

For this set of reasons it is useful to apply the path integral techniques developed in \cite{Dudal:2020yah}
to a more complicated situation by incorporating a chiral medium, in light of eventually moving towards \((1+1)\)d \(\mathbb{C}P^{N-1}\) or \((2+1)\)d Yang-Mills theories.
Moreover it is interesting what the impact of a general set of boundary conditions and a chiral medium is on the 3D effective boundary theory resulting from integrating over the photons. We will pay particular attention to imposing these boundary conditions in a gauge invariant manner. The presented methodology is quite flexible, and where possible, we will check against existing results.

\section{Setup}

Consider the Euclidean QED action, augmented with a $\theta$ term
\begin{equation}\label{eq:axion-action}
    S = \int \dd[4]{x} \qty[\frac{1}{4} F_{\mu\nu} F_{\mu\nu} + \frac{i}{4}g\theta F_{\mu\nu} \dual{F}_{\mu\nu} ] \;,
\end{equation}
where \(F_{\mu\nu}\) is the Maxwell field strength tensor
\begin{equation}
    F_{\mu\nu} = \partial_\mu A_\nu - \partial_\nu A_\mu \;,
\end{equation}
and \(\dual{F}_{\mu\nu}\) is the dual field strength
\begin{equation}
    \dual{F}_{\mu\nu} = \frac{1}{2} \varepsilon_{\mu\nu\rho\sigma} F_{\rho\sigma} \;.
\end{equation}
It is worth emphasizing that the action above is not axion QED \cite{Wilczek:1987mv}.
In the present case, the field $\theta$ is a fixed classical background field describing a chiral material.
The term proportional to $\theta$ breaks {\sl CP}-invariance, such that the no-go theorem for a repulsive Casimir force \cite{Kenneth:2006vr} can be circumvented.
Such a term can be used to model a variety of chiral materials such as topological insulators \cite{Qi:2008ew} and Weyl semimetals \cite{Grushin:2012mt}.
In short, a time reversal invariant topological insulator can be modeled by\footnote{The theta term is more conventionally written as \(\frac{\theta^\prime}{16\pi^2} F_{\mu\nu}\tilde{F}_{\mu\nu}\), in which case a topological insulator corresponds to \(\theta^\prime = \pi\).} \(g\theta=\frac{1}{4\pi}\) while a Weyl semimetal corresponds to a linear \(\theta\).
In general, the energy distance between two Weyl nodes corresponds to a time-like component $b_0$, and the momentum distance to a space-like one $\vec b$. Usually, $b_\mu\propto \partial_\mu \theta$, see  \cite{Goswami:2012db} for a derivation of an effective field theory, see also \cite{Chernodub:2013kya}.
A similar theory was considered in \cite{Carroll:1989vb} in a different context.

There exist a variety of configurations which are of interest for the Casimir effect,
often consisting of vacuum or some other material of interest between slabs of topological insulators \cite{Grushin:2010qoi} or Weyl semimetals \cite{Farias:2020qqp,Wilson:2015wsa,Jiang:2018ivv}.
Our setup is more in line with \cite{Fukushima:2019sjn,Kharlanov:2009pv}, where a chiral material, modeled with a linear \(\theta\), confined between infinite slabs of a perfectly conducting material.
Finally it should be noted that in all these setups a repulsive Casimir force can be achieved, signaling that the no-go theorem has indeed been circumvented,
except in \cite{Kharlanov:2009pv} where the gradient of \(\theta\) is chosen to be time-like. Notice that this case is also what could be relevant for the quark-gluon plasma where a chiral imbalance is modelled via a chiral chemical potential $\propto \partial_t\theta$ \cite{Kharzeev:2013ffa}.

Our setup consists of two infinite parallel plates normal to the \(z\)-axis at positions \(z=\pm \frac{L}{2}\).
The boundary conditions enforced on these plates are elaborated on more in the next section.
Between the parallel plates we place a chiral medium characterized by \(\partial_z \theta = \beta\) with \(\beta\) a constant.
This can be seen as incorporating a chiral medium in a way that minimally breaks Lorentz symmetry, as introducing the chiral medium does not break the Lorentz symmetry of the vacuum any more than the plates already have.
If the gradient of \(\theta\) did not lie along the \(z\)-axis, then gauge invariance would be broken, this is however a so-called consistent gauge anomaly as the chiral anomaly of fermionic surface states causes the total action to be gauge invariant \cite{Goswami:2012db,Qi:2008ew,Callan:1984sa}.
This setup is similar to the one used in \cite{Fukushima:2019sjn} (although the boundary conditions on the plates have been left unspecified so far),
with the
biggest difference being that in our case the plates are infinitely thin, allowing the fields to propagate outside of the plates.

\begin{figure}
    \centering
    \begin{equation*}
        \begin{tikzpicture}[scale=.75,baseline=0]
            \fill[gray] (-2,-3) rectangle (2,3);
            \draw (-5,-2) -- (-2,-2) -- (2,2) -- (5,2) node[above right] {\(\theta(z)\)};
            \node[align=center] at (0, 2) {Chiral QED/\\Weyl Semimetal};
            \node at (0,-1.5) {\(\theta(z) \propto z \beta \)};
            \node at (-3.5, 1.5) {QED};
            \node at (-3.5,-1.5) {\(\theta(z) = \) const};
            \node at (3.5, 1.5) {QED};
            \node at (3.5,-1.5) {\(\theta(z) = \) const};
            \draw[dashed] (-2,-2) -- (-2,-3.2) node[below] {\(z=-\frac{L}{2}\)};
            \draw[dashed] (2,2) -- (2,-3.2) node[below] {\(z=\frac{L}{2}\)};
            \draw[-latex, thin] (-5,0) -- (5,0) node[right] {\(z\)};
        \end{tikzpicture}
        +\quad
        \begin{tikzpicture}[scale=.75,baseline=0]
            \draw[thick] (-2,3) node[left] {\(\theta_-\)} -- (-2,-3) node[below] {\(z=-\frac{L}{2}\)};
            \draw[thick] (2,3) node[right] {\(\theta_+\)} -- (2,-3) node[below] {\(z=\frac{L}{2}\)};
            \draw[-latex, thin] (-5,0) -- (5,0) node[right] {\(z\)};
        \end{tikzpicture}
    \end{equation*}
    \caption{Our setup, consisting of two parts.
    On the left the profile of the \(\theta\) background field is shown, which divides the vacuum into parts where conventional QED and chiral QED, describing a Weyl semimetal, apply.
    On the right the thin parallel plates are shown, on which PEMC boundary conditions are applied with parameters \(\theta_\pm\).}
    \label{fig:theta}
\end{figure}
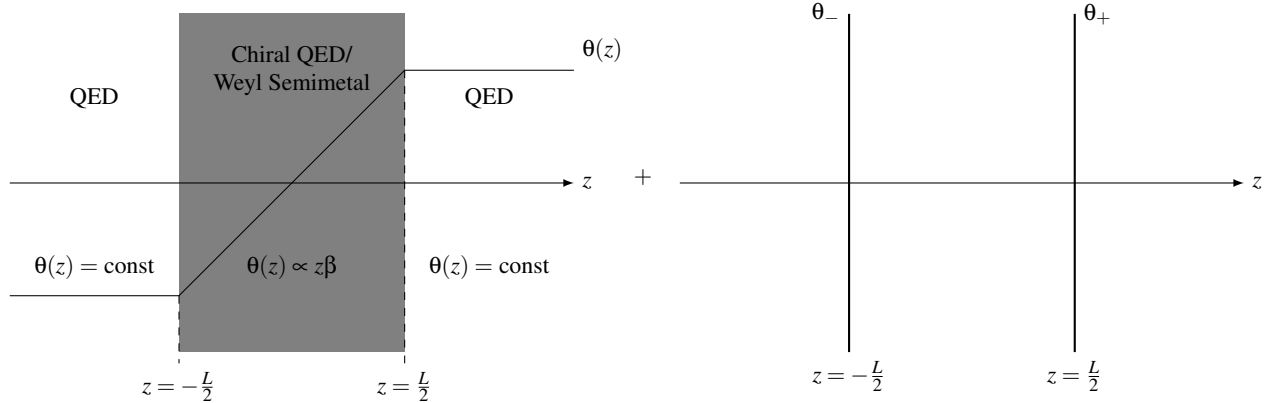

The behaviour of \(\theta\) outside of the plates needs some extra care.
A jump in the value of \(\theta(z)\) induces Chern-Simons terms on the surface of the discontinuity \cite{Canfora:2011fd}.
In other words, this causes Hall currents to exist on the surface.
While these ``Chern-Simons'' surfaces are also of interest for repulsive Casimir effects \cite{Fialkovsky:2018fpo} we want to avoid discontinuities in \(\theta\) as they can cause the electric and magnetic fields, which are components of \(F_{\mu\nu}\), to be discontinuous.
The background field \(\theta\) can consequently be written down as
\begin{equation}\label{eq:fulltheta}
    \theta(z) = \qty[H\qty(z+\frac{L}{2})\qty(z+\frac{L}{2}) - H\qty(z-\frac{L}{2}) \qty(z-\frac{L}{2}) - \frac{L}{2}] \beta + \theta_0 \qty(\beta\;L) \;,
\end{equation}
where \(H(z)\) is the Heaviside step function, and \(\theta_0\) is an arbitrary function of \(\beta\;L\) which does not affect the physics, courtesy of the fact that only the gradient of \(\theta\) enters the equations of motion.
The behaviour of \(\theta(z)\) is shown in Figure \ref{fig:theta}.
It follows that, in a distributional sense,
\begin{equation}
    \partial_z \theta(z) = \beta(z) = \qty[H\qty(z+\frac{L}{2}) - H\qty(z-\frac{L}{2})] \beta \;,
\end{equation}
where we make the distinction between \(\beta(z)\) as a function of \(z\) and the constant value \(\beta\) between the plates, such that \(\partial_z \theta(z) = \beta\) when \(-\frac{L}{2} < z < \frac{L}{2}\).

We also compare this setup with the case where \(\partial_z \theta(z) = \beta\) on the whole space, i.e.\ when the entire vacuum is chiral, and the QED case where \(\beta=0\).

\section{Boundary conditions and gauge fixing}

From the variational principle a set of natural boundary conditions can be derived,
for this reason we first place the model on a manifold \(\mathcal{M}\) with boundary \(\partial\mathcal{M}\).
Varying the action \eqref{eq:axion-action} with respect to $A_\mu$, but keeping the boundary terms arising from integration by parts, results in
\begin{equation}\label{eq:action-variation}
    \delta_A S = \int_\mathcal{M} \var{A_\mu} \qty[\partial_\nu F_{\mu\nu} + ig \partial_\nu \theta \dual{F}_{\mu\nu}] + \int_{\partial \mathcal{M}} \var{A_\mu} n_\nu \qty[F_{\mu\nu} + ig \theta \dual{F}_{\mu\nu}]
\end{equation}
which when requiring that the action is stationary become the equations of motion with boundary conditions
\begin{equation}\label{eq:organic-boundary-conditions}
\begin{aligned}
    \partial_\nu F_{\mu\nu} + ig \partial_\nu \theta \dual{F}_{\mu\nu} &= 0 \;, \\
    \eval{n_\nu F_{\mu\nu} + ig \theta n_\nu \dual{F}_{\mu\nu}}_{\partial \mathcal{M}} &= 0 \;,
\end{aligned}
\end{equation}
where \(n_\mu\) is the normal vector of the boundary. At first sight, there is another way of implementing boundary conditions consistent with the variational principle, namely fixing the value of \(A_\mu\) on \(\partial\mathcal{M}\) and setting the variation \(\var{A_\mu}\) zero on the boundary. Unfortunately these are not necessarily gauge invariant and so we do not consider these. For the record, we notice here that \cite{Fukushima:2019sjn} choose to set $A_\mu\equiv0$ on the boundary.  A thorough discussion on boundary conditions for gauge theories, consistent with the gauge principle, can be found in \cite{Vassilevich:2003xt}. It can be checked our implementation can be reformulated in terms of those of \cite{Vassilevich:2003xt}.

The gauge invariant boundary conditions correspond to a mix between perfectly electric conducting (PEC) \(n_\mu \dual{F}_{\mu\nu}=0\) and perfectly magnetic conducting (PMC) \(n_\mu F_{\mu\nu}=0\),
dependent on the value of $\theta$ on the boundary.
These boundary conditions are known in literature as ``perfect electromagnetic boundary conditions'' (PEMC) \cite{lindell2005perfect} or ``Chern-Simons boundary conditions'' \cite{Bordag:1999ux}, and they appear naturally in the description of chiral optical or bi-isotropic materials \cite{lindell1994electromagnetic}.
Chiral metamaterials were already explored in e.g.~\cite{Zhao:2009zz,Silveirinha:2010zz,Zhao:2010zzq} in the light of a repulsive Casimir force.

In our case the manifold is flat Euclidean space and the fields are assumed to decay sufficiently quickly such that these boundary conditions can be ignored,
although it is still interesting to use these boundary conditions in the parallel plate setup for the Casimir effect.

As an example consider standard QED where \(\theta=0\), such that the natural boundary conditions become PMC boundary conditions
\begin{equation}
    n_\mu F_{\mu\nu} = 0 \;.
\end{equation}
These boundary conditions can be seen as the dual of the PEC boundary conditions,
\begin{equation}
    n_\mu \dual{F}_{\mu\nu} = 0 \;,
\end{equation}
which are conventionally chosen as the boundary conditions on the plates for the Casimir effect.
PEC boundary conditions correspond to perfectly conducting plates, i.e.\ having zero resistance,
which is also why PMC boundary conditions are sometimes called dual superconductor boundary conditions (DSBC).
These boundary conditions are the Abelian analog of the ``bag'' boundary conditions used when modeling nucleons
as a bubble stabilized by the Casimir force \cite{Bordag:2001qi,Plunien:1986ca,Hasenfratz:1977dt}.
While the ``bag'' boundary conditions are derived from assuming that the QCD vacuum is a perfect color dia-electric \cite{Lee:1981mf},
it follows that assuming that there are no fields outside of the ``bag'' at least produces the same results,
although no fermions have been taken into account.

With that in mind we choose to enforce boundary conditions inspired by \eqref{eq:organic-boundary-conditions}
on the plates, namely
\begin{equation}\label{eq:actual-bc}
    n_\nu F_{\mu\nu} + ig \theta_\pm n_\nu \tilde{F}_{\mu\nu} \eval{}_{z=\pm\frac{L}{2}} = 0 \;,
\end{equation}
where \(\theta_\pm\) are constant and real parameters.
To simplify our notation we work with the boundary conditions \eqref{eq:organic-boundary-conditions}, and at the end of the calculations, replace \(\theta(z=\pm\frac{L}{2}) \to \theta_\pm\).
It should be stressed that this does not mean we modify the \(\theta(z)\) background field at \(z=\pm\frac{L}{2}\), it is simply abuse of notation and \(\theta_\pm\) remain free material parameters.
Furthermore \(\theta(z)\) will only appear explicitely in calculations regarding the boundary conditions, as in the bulk only \(\partial_z \theta(z)\) is relevant as can be seen in \eqref{eq:action-variation}.
Consequently this way of dealing with the boundary conditions does not interfere with the \(\theta\) background.

Following \cite{Bordag:1983zk,Dudal:2020yah} the boundary conditions on a surface \(\Sigma\) can be applied in a minimal way by representing the functional delta in terms of an auxiliary field
\begin{equation}
    \delta_\Sigma[f_\mu] = \int \DD{b_\mu} e^{-\int_\Sigma \dd[3]{\vec{x}} b_\mu f_\mu } \;.
\end{equation}
Applying this to the parallel plates at \(z=\frac{L}{2}\) (\(\Sigma\)) and \(z=-\frac{L}{2}\) (\(\bar{\Sigma}\)) the action can be written as
\begin{equation}
\begin{aligned}
    S &= \int_\mathcal{M} \qty[\frac{1}{4} F_{\mu\nu} F_{\mu\nu} + \frac{i}{4} g \theta F_{\mu\nu} \dual{F}_{\mu\nu}] + \frac{1}{2}\int_{\Sigma} b_\mu n_\nu \qty[F_{\mu\nu} + ig \theta_+ \dual{F}_{\mu\nu}] + \frac{1}{2} \int_{\bar{\Sigma}} \bar{b}_{\mu} n_\nu \qty[F_{\mu\nu} + ig \theta_- \dual{F}_{\mu\nu}] \\
    &= \int\dd[4]{x} \qty[\frac{1}{4} F_{\mu\nu} F_{\mu\nu} + \frac{i}{4} g \theta F_{\mu\nu} \dual{F}_{\mu\nu} + \frac{1}{2}\qty(\delta\qty(z-\frac{L}{2}) b_\mu + \delta\qty(z+\frac{L}{2})\bar{b}_\mu)n_\nu\qty(F_{\mu\nu} + i g \theta \dual{F}_{\mu\nu})] \;,
\end{aligned}
\end{equation}
where \(n_\mu=(0,0,0,1)\) is the unit vector normal to the plates and on the last line we abused our notation to replace the \(\theta_\pm\) with \(\theta(z)\).
Remark that due to the antisymmetry of \(F_{\mu\nu}\) (\(\dual{F}_{\mu\nu}\)) the action is independent of the value of \(b_3\) (\(\bar{b}_3\)).
This is a sign that \(b_\mu\) and \(\bar{b}_\mu\) are not four-vectors restricted to the hypersurfaces \(\Sigma\) and \(\bar{\Sigma}\), but can be seen as source terms of the electromagnetic potential ``native'' to \(\Sigma\) and \(\bar{\Sigma}\).
This allows us to immediately integrate over \(b_3\) and \(\bar{b}_3\) as this results in an irrelevant constant.
Using the standard BRST symmetry
\begin{equation}
    sA_\mu = \partial_\mu c \qc sc^\dag = h \qc sc = 0 \qc sh = 0 \;,
\end{equation}
a gauge fixing term can be introduced via
\begin{equation}
\begin{aligned}
    S_\text{gf} &= s \int \dd[4]{x} \qty[ \frac{\xi}{2} h c^\dag + c^\dag \partial_\mu A_\mu] \\
    &= \int \dd[4]{x} \qty[\frac{\xi}{2}h^2 + h\partial_\mu A_\mu - c^\dag \partial^2 c] \;,
\end{aligned}
\end{equation}
where, as is usual in QED, the ghosts are decoupled.
The ghosts will consequently not contribute to the theory and can be integrated out for convenience.
In a similar fashion, the Lagrange multiplier \(h\) will complicate the propagator calculation, and can be integrated out by completing the square.
This essentially sets \(h=-\frac{1}{\xi} \partial_\mu A_\mu\) such that the final gauge fixing term is
\begin{equation}
    S_\text{gf} = \int \dd[4]{x} \qty[-\frac{1}{2\xi} \qty(\partial A)^2] \;.
\end{equation}
After integration by parts the action can be rewritten as
\begin{equation}\label{eq:starting-point-action}
    S = - \frac{1}{2} \int \dd[4]{x} \qty[ A_\mu K_{\mu\nu}^\xi A_\nu + \qty(\delta\qty(z-\frac{L}{2}) b_\mu + \delta\qty(z+\frac{L}{2})\bar{b}_\mu)n_\nu\qty(F_{\mu\nu} + i g \theta \dual{F}_{\mu\nu})] \;,
\end{equation}
where the kinetic operator is given by
\begin{equation}\label{eq:kinetic-pos}
    K_{\mu\nu}^\xi = \delta_{\mu\nu} \partial^2 - \qty(1-\frac{1}{\xi}) \partial_\mu \partial_\nu + ig \varepsilon_{\mu\nu\rho 3} \beta(z) \partial_\rho \;.
\end{equation}
The action \eqref{eq:starting-point-action} is the starting point for our calculation of the Casimir force,
and from here on out we choose the Feynman gauge \(\xi=1\) and write \(K_{\mu\nu} \equiv K^{\xi=1}_{\mu\nu}\) to simplify calculations.

\section{Reduction to a 3D effective theory}

Now that the boundary conditions are incorporated directly in the action with the help of auxiliary fields, one can in principle integrate them out.
This has been done in \cite{Bordag:1983zk} for the QED case, and results in QED with a modified photon propagator that respects the boundary conditions.
Instead of this we take a closer look at the auxiliary fields.
The way the auxiliary fields act as sources confined to the plate geometry is reminiscent of so-called defect QFT \cite{Grignani:2019zxc},
although in this case the field does not propagate.
We can, however, choose to integrate out the photon and, as the photon is the only field that extended in the \(z\)-direction, integrate explicitly over \(z\).
This would result in a non-local dimensionally reduced effective theory for the ``defect'' fields \(b_\mu\) and \(\bar{b}_\mu\). 
This is what marks the flexibility of our computational framework.
Unlike in other Casimir energy/force computations, we do not need to solve for the full Green's function obeying the boundary conditions.
Rather we separately focus on the the ``pre-Green's'' function (inverse of $K_{\mu\nu}^\xi$), after which, if present, the influence of the boundary conditions on the plates can be added as encoded in the $(b_\mu, \bar{b}_\mu)$-dynamics.

In section \ref{sec:polarization} we perform a Fourier transform over the directions normal to the \(z\)-axis, but keep the \(z\)-coordinate.
This partially avoids the complexities arising from the (partially) broken translation invariance, which arises from the chiral medium which is only present between the plates.
Furthermore a polarization basis is introduced in which the Lorentz structure of the kinetic operator is diagonal.
In this ``mixed Fourier-real'' space we calculate the photon Green's function in section \ref{sec:greenfunction},
where the main difficulty follows from gluing the solutions together from the chiral QED and normal QED regions.
Finally we integrate out the photon and construct the effective theory in section \ref{sec:effective-theory}.

\subsection{Partial Fourier transformation and polarization basis}\label{sec:polarization}

A first step to calculate the Green's function of a field is often to perform a Fourier transformation, which translates the differential equation
\begin{equation}
    K_{\mu\rho} K^{-1}_{\rho\nu}(x-y) = \delta_{\mu\nu} \delta^{(4)}\qty(x-y) \;,
\end{equation}
to an algebraic one.
A problem arises from the fact that \(\partial_\mu \theta\) depends on the spacetime coordinate, and is essentially a block function taking the value \(\beta\) when \(-\frac{L}{2} < z < \frac{L}{2}\) and zero otherwise.
It follows that Fourier transforming the \(z\) coordinate would involve a convolution in momentum space.
For this reason we only Fourier transform the \(t,x,y\) coordinates, after which the action becomes
\begin{equation}\label{eq:fourier-action}
    S = \frac{1}{2} \int \frac{\dd[3]{\vec{k}}}{\qty(2\pi)^2} \dd{z} \qty[A_\mu(\vec{k},z) K_{\mu\nu}\qty(\vec{k},\partial_z^2) A_\nu\qty(-\vec{k},z) + \qty(B_\mu(\vec{k},z) + \bar{B}_\mu(\vec{k},z)) A_\mu(-\vec{k},z)] \;,
\end{equation}
where the kinetic operator is now
\begin{equation}\label{eq:kin-cbasis}
    K_{\mu\nu}\qty(\vec{k},\partial_z^2) = \delta_{\mu\nu} \qty(\partial_z^2 - \vec{k}^2) - g \beta(z) \epsilon_{\mu\nu j 3} k_j \;,
\end{equation}
and the terms arising from the boundary conditions have been written as sources
\begin{equation}
    \begin{aligned}
        B_\mu(\vec{k},z) &= b_i(\vec{k})V_{i\mu}(\vec{k},z)\delta\qty(z-\frac{L}{2}) + b_i\qty(\vec{k})\delta^\prime\qty(z-\frac{L}{2}) \delta_{i \mu} \;, \\
        \bar{B}_\mu(\vec{k},z) &= \bar{b}_i(\vec{k})V_{i\mu}(\vec{k},z)\delta\qty(z+\frac{L}{2}) + \bar{b}_i\qty(\vec{k})\delta^\prime\qty(z+\frac{L}{2}) \delta_{i \mu}  \;, \\
        V_{i\mu}(\vec{k},z) &= ik_i \delta_{3\mu} + g\theta(z)\varepsilon_{ijk} k_k \delta_{j \mu} \;.
    \end{aligned}
\end{equation}
The latin indices \(i,j,k,\cdots\) take the values \(0,1,2\) corresponding to the \(t,x,y\) axes,
and denote the directions normal to the \(z\)-axis.
The Green's function is now the solution of the second order differential equation
\begin{equation}\label{eq:green-eq}
    K_{\mu\rho}(\vec{k},\partial_z^2) K_{\rho\nu}^{-1}\qty(\vec{k},z,z^\prime) = \delta_{\mu\nu} \delta(z-z^\prime) \;.
\end{equation}
This equation simplifies significantly when instead of using the coordinate basis, a new polarization basis is chosen.
We introduce the real (linear) polarization vectors \(E^0_\mu, E^1_\mu, E^2_\mu, E^3_\mu\), and denote components with respect to a polarization basis with the latin indices \(r,s,t,\cdots \in \{0,1,2,3\}\).
As usual repeated indices correspond to a summation.
These vectors should form an orthonormal basis
\begin{equation}
    E_\mu^r {E_\mu^s}^\dag = \delta^{rs} \qc E_\mu^r {E^r_\nu}^\dag = \delta_{\mu\nu} \;,
\end{equation}
and from the structure of \eqref{eq:kin-cbasis} it follows that two obvious choices are
\begin{equation}
    E^0_i = \frac{k^{i}}{\abs{\vec{k}}} \qc E^0_3=0 \qc E^3_i = 0 \qc E^3_3 = 1 \;,
\end{equation}
such that \(E^0\) is the longitudinal polarization%
\footnote{Technically it is the polarization in the direction of \(\vec{k}\), which is normal to the \(z\)-axis, and is not the usual definition of the longitudinal polarization as lying in the direction of the spatial part of the four-momentum.} and \(E^3\) is the polarization in the \(z\) direction\footnote{This polarization is then analogous to timelike polarization.}.
The two polarizations that are left should obey
\begin{equation}
    E^r_i k_i = 0\qc E^r_3=0 \qq{for} r=1,2 \;,
\end{equation}
and are consequently the two $3D$ transversal polarizations i.e.\ \(\qty(\delta_{ij} - \frac{k_i k_j}{\vec{k}^2})E_j^r=E_i^r\) for \(r=1,2\).
As the two polarizations should be mutually orthogonal it follows that one can choose
\begin{equation}
    E^2_i = \varepsilon_{ijk} \frac{k_k}{\abs{\vec{k}}} E_j^1
\end{equation}
such that given a real transversal polarization vector \(E_i^1\) with \(E^1_i E^1_i=1\) we also have a \(E^2_i\) with \(E^2_i E^2_i = 1\) and \(E^2_i E^1_i = 0\). Inverting gives
\begin{equation}
    \varepsilon_{ijk} \frac{k_k}{\abs{\vec{k}}} E_j^2 = - E_i^1 \;.
\end{equation}
In this basis the \(\epsilon_{\mu\nu j 3}\) term of \eqref{eq:kin-cbasis} becomes
\begin{equation}
    E_\mu^r\qty(\vec{k}) g \beta(z) \epsilon_{\mu\nu j 3} k_j (E_\nu^s\qty(\vec{k}))^{\dagger} = \mqty( 0 & 0 & 0 & 0\\ 0 & 0 & -g\beta(z) \abs{\vec{k}} & 0 \\ 0 & g\beta(z) \abs{\vec{k}} & 0 & 0 \\ 0 & 0 & 0 & 0) \;.
\end{equation}
This can be simplified further by using different basis vectors \(\tilde{E}^1_\mu, \tilde{E}^2_\mu\) constructed from a complex linear combination of the \(E^1_\mu, E^2_\mu\) polarizations%
\footnote{These polarizations are analogous to circular polarizations.}%
\begin{equation}
    \tilde{E}^0_\mu = E^0_\mu \qc \tilde{E}^1_i =\frac{1}{\sqrt{2}}\qty(E^1_i + i E^2_i) \qc \tilde{E}^2_i = \frac{1}{\sqrt{2}} \qty(E^1_i - i E^2_i) \qc \tilde{E}^3_\mu = E^3_\mu
\end{equation}
In this basis it follows that \(\epsilon_{ijk}k_k\) is diagonal
\begin{equation}
    \tilde{E}^r_\mu g\beta(z)\varepsilon_{\mu \nu j 3} k_j \qty(\tilde{E}_\nu^s)^\dag = \diag(0 \qc +ig\beta(z)\abs{\vec{k}} \qc - ig\beta(z)\abs{\vec{k}} \qc 0 ) \;,
\end{equation}
such that the kinetic operator has been reduced to a diagonal matrix of the form
\begin{equation}\label{eq:kin-ebasis}
    K_{rs}\qty(\vec{k},\partial_z^2) = \diag(\partial_z^2 - \abs{\vec{k}}^2\qc \partial_z^2 - \qty(k_c^\star)^2(z)\qc \partial_z^2 - k_c^2(z)\qc \partial_z^2 - \abs{\vec{k}}^2) \;,
\end{equation}
where \(k_c^2(z) = \abs{\vec{k}}^2 + ig\beta(z) \abs{\vec{k}}\) has been defined.
As can be seen, only the ``transversal'' polarizations \({r=s=1,2}\) depend on the parameter \(\beta(z) \equiv \partial_z \theta(z)\),
consequently only these polarizations require extra care when constructing the Green's function.

\subsection{The Green's function in mixed Fourier space}\label{sec:greenfunction}

Using the \(\tilde{E}\) basis constructed in the previous section we have that the differential equation for the Green's function \eqref{eq:green-eq} becomes
\begin{equation}
    K_{rt}(\vec{k},\partial_z^2)K^{-1}_{ts}(\vec{k},z,z^\prime) = \delta_{rs} \delta(z-z^\prime) \;,
\end{equation}
and using the explicit form of \(K_{rt}\) in this basis \eqref{eq:kin-ebasis} it follows that \(K^{-1}_{rs}\) only consists of two functions
\begin{equation}\label{eq:prop-components}
    \begin{aligned}
        \qty(\partial_z^2 - \abs{\vec{k}}^2) f(\vec{k},z-z^\prime) &= \delta(z-z^\prime) \;, \\
        \qty(\partial_z^2 - k_c^2(z)) D(\vec{k},z,z^\prime) &= \delta(z-z^\prime) \;,
    \end{aligned}
    \begin{aligned}
        K_{00}^{-1} = K_{33}^{-1} &= f(\vec{k},z-z^\prime) \;, \\
        K_{22}^{-1} = \qty(K_{11}^{-1})^\star &= D(\vec{k},z,z^\prime) \;.
    \end{aligned}
\end{equation}
To make the discussion easier it is useful to take the \(z\) dependence out of \(k_c^2(z)\)
\begin{equation}
    k_c^2(z) = \left\lbrace\begin{aligned}
        k_c^2 = \abs{\vec{k}}^2 + ig\beta \abs{\vec{k}} &\qq{if} z \in \qty[-\frac{L}{2},\frac{L}{2}] \;, \\
        \abs{\vec{k}}^2 &\qq{if} z \notin \qty[-\frac{L}{2},\frac{L}{2}] \;,
    \end{aligned}\right.
\end{equation}
where \(k_c^2\) and \(\beta\) are meant as constants, i.e.\ their value within the plates.
The last equation can be written as
\begin{equation}\label{def_k_cz}
k_{c}^{2}(z) =  \abs{\vec{k}}^2 + ig\beta \abs{\vec{k}}\qty(H\qty(z+\frac{L}{2}) - H\qty(z- \frac{L}{2})) \;.
\end{equation}
With this decomposition it can be seen that \(D(\vec{k},z,z^\prime)\) solves two differential equations depending on the region
\begin{equation}
    \begin{aligned}
        \qty(\partial_z^2 - \abs{\vec{k}}^2) D(\vec{k},z,z^\prime) &= \delta(z-z^\prime) &\qq{if} z \in \qty[-\frac{L}{2},\frac{L}{2}]  \;, \\
        \qty(\partial_z^2 - k_c^2) D(\vec{k},z,z^\prime) &= \delta(z-z^\prime) &\qq{if} z \notin \qty[-\frac{L}{2},\frac{L}{2}] \;.
    \end{aligned}
\end{equation}
Consequently \(D(\vec{k},z,z^\prime)\) can be constructed from the basic Green's functions
\begin{equation}
    \begin{aligned}
        \qty(\partial_z^2 - \abs{\vec{k}}^2) f(\vec{k},z-z^\prime) &= \delta(z-z^\prime) \;, \\
        \qty(\partial_z^2 - k_c^2) \varphi(\vec{k},z-z^\prime) &= \delta(z-z^\prime) \;,
    \end{aligned}
\end{equation}
where \(f(\vec{k},z-z^\prime)\) coincidentally is also the solution for \(K_{00}\) and \(K_{33}\).
The complete solution \(D(\vec{k},z,z^\prime)\) then consists of the basic Green's functions
\begin{equation}
    f(\vec{k},z-z^\prime) = -\frac{1}{2\abs{\vec{k}}} e^{-\abs{z-z^\prime}\abs{\vec{k}}} \qc \varphi(\vec{k},z-z^\prime) = -\frac{1}{2k_c}  e^{-\abs{z-z^\prime}k_c} \;,
\end{equation}
and the solutions to the homogeneous equations
\begin{equation}
    \begin{aligned}
        \qty(\partial_z^2 - \abs{\vec{k}}^2)f_0 &= 0 \qc f_0 \in \{e^{\abs{\vec{k}} z}, e^{-\abs{\vec{k}} z}\} \;, \\
        \qty(\partial_z^2 - k_c^2) \varphi_0 &= 0 \qc \varphi_0 \in \{\sinh(k_c z), \cosh(k_c z)\} \;.
    \end{aligned}
\end{equation}
The homogeneous solutions \(f_0\) have been written in terms of exponentials, as they are the solutions in the conventional QED medium, and requiring that the fields decay to zero at infinity naturally picks out one of the exponentials on either side of the plates.
The solutions \(\varphi_0\) do not have any such restrictions, as the chiral medium is confined between the two QED media, and hyperbolic functions are the most convenient expression to use in this case.
The complete solution \(D(\vec{k},z,z^\prime)\) can be constructed by splitting it into three parts
\begin{equation}
    D(\vec{k},z,z^\prime) = \left\lbrace\begin{aligned}
        D^+ &\qif z > \frac{L}{2} \;, \\
        \bar{D} &\qif -\frac{L}{2} < z < \frac{L}{2} \;,\\
        D^- &\qif z < -\frac{L}{2} \;,
    \end{aligned}\right.
\end{equation}
after which the general solution in each medium can be written down as
\begin{equation}
    \begin{aligned}
        D^\pm(\vec{k},z,z^\prime) &= C^\pm(z^\prime) e^{-\abs{\vec{k}} \abs{z}} + f(\vec{k},z-z^\prime) \;, \\
        \bar{D}\qty(\vec{k},z,z^\prime) &= C_1(z^\prime) \cosh(k_c z) + C_2(z^\prime) \sinh(k_c) + \varphi(\vec{k},z-z^\prime) \;,
    \end{aligned}
\end{equation}
and \(C_1(z'), C_2(z'), C^\pm(z')\) are four complex functions that should be chosen such that \(D(\vec{k},z,z^\prime)\) is continuous and smooth.
Explicitly this condition requires that the solutions outside of the chiral medium should be properly glued to the inner solution
\begin{equation}
    \begin{aligned}
        \eval{D^\pm(\vec{k})\qty(\vec{k},z,z^\prime)}_{z=\pm\frac{L}{2}} &= \eval{\bar{D}(\vec{k},z,z^\prime)}_{z=\pm\frac{L}{2}} \;, \\
        \eval{\partial_z D^\pm(\vec{k})\qty(\vec{k},z,z^\prime)}_{z=\pm\frac{L}{2}} &= \eval{\partial_z \bar{D}(\vec{k},z,z^\prime)}_{z=\pm\frac{L}{2}} \;,
    \end{aligned}
\end{equation}
which enforces continuity and smoothness.
It is useful to define the following constants for writing down the explicit form of \(C_1(z'), C_2(z')\) and \(C^\pm(z')\)
\begin{equation}
    \begin{aligned}
        \gamma^\pm &= \qty(1\pm \frac{k_c}{\abs{\vec{k}}}) \;, \\
        N^\pm &= \gamma^+ e^{\frac{L}{2}k_c} \pm \gamma^- e^{-\frac{L}{2}k_c} \;.
    \end{aligned}
\end{equation}
The explicit form of \(C_1(z'), C_2(z')\) and \(C^\pm(z')\) depends on whether \(z^\prime\) lies inside or outside of the chiral medium,
and are written down in Appendix \ref{apx:green-coef}.
The part that will be relevant to the calculation of the effective boundary theory in section \ref{sec:effective-theory} is when \(z,z^\prime \in \qty[-\frac{L}{2},\frac{L}{2}]\),
and is given by
\begin{equation}\label{eq:middle-prop}
    \bar{D}(\vec{k},z,z^\prime) = \frac{\gamma^-}{N^+ k_c} e^{-k_c \frac{L}{2}} \cosh(k_c z) \cosh(k_c z^\prime) + \frac{\gamma^-}{N^- k_c} e^{-k_c \frac{L}{2}} \sinh(k_c z) \sinh(k_c z^\prime) + \varphi(z-z^\prime) \;.
\end{equation}

It can be shown that \(D(\vec{k},z,z^\prime)\) obeys
\begin{equation}
    D(\vec{k},z,z^\prime) = D(\vec{k},-z,-z^\prime) = D(\vec{k},z^\prime,z) \;,
\end{equation}
and similarly for \(f(\vec{k},z-z^\prime)\).
The polarization vectors \(\tilde{E}^r_\mu\) do not depend on \(z\), and so the propagator in the coordinate basis \(K_{\mu\nu}^{-1}(\vec{k},z,z^\prime)\) has those same properties.

Using \eqref{eq:prop-components} it is now possible to construct the Green's function.
It should be remarked that this is not the full photon propagator of the theory,
as the boundary conditions have not yet been taken into account.
If we would integrate over the \(b,\bar{b}\) auxiliary fields such as in \cite{Bordag:1983zk}
the propagator would be modified to obey the boundary conditions.
In a theory with a translation invariant vacuum this would break the translation invariance of the propagator.
We will do the opposite however, and integrate the photon field \(A_\mu\) out to arrive at an effective 3D boundary theory as in \cite{Dudal:2020yah}.

\subsection{The effective boundary theory}\label{sec:effective-theory}

Now that \(K_{\mu\nu}^{-1}\) has been calculated we can perform the shift of variables
\begin{equation}
    A_\mu(\vec{k},z) \to A_\mu(\vec{k},z) - \frac{1}{2} \int \dd{z^\prime} K_{\mu\nu}^{-1}(\vec{k},z,z^\prime) \qty(B_\nu(\vec{k},z^\prime) + \bar{B}_{\nu}(\vec{k},z^\prime)) \;,
\end{equation}
which leaves the measure invariant \(\DD{A^\prime} \DD{b} \DD{\bar{b}} = \DD{A} \DD{b} \DD{\bar{b}}\).
This shift brings the action (\ref{eq:fourier-action}) into the form \(S = S_A + S_{b\bar{b}}\) with
\begin{equation}\label{eq:shifted-action}
    \begin{aligned}
        S_{b\bar{b}} &= -\frac{1}{4} \int \frac{\dd[3]{\vec{k}}}{\qty(2\pi)^3} \int \dd{z} \int \dd{z^\prime} \qty[\qty(B_{\mu}(\vec{k},z) + \bar{B}_\mu(\vec{k},z)) K^{-1}_{\mu\nu}(\vec{k},z,z^\prime) \qty(B_\nu(-\vec{k},z^\prime) + \bar{B}_\nu(-\vec{k},z^\prime))] \\
        S_A &= \frac{1}{2} \int \frac{\dd[3]{\vec{k}}}{\qty(2\pi)^3} \int \dd{z} A_\mu(\vec{k},z) K_{\mu\nu}\qty(\vec{k},\partial_z^2) \;. A_\mu\qty(-\vec{k},z)
    \end{aligned}
\end{equation}
Consequently we can now integrate over the \(A_\mu\) field to obtain
\begin{equation}\label{eq:A-integration}
    \int \DD{A} \DD{b} \DD{\bar{b}} e^{-S_A - S_{b\bar{b}}} = \frac{C}{\sqrt{\det(K)}} \int \DD{b}\DD{\bar{b}}e^{-S_{b\bar{b}}} \;,
\end{equation}
with \(C\) an irrelevant constant.
If no extra boundary conditions would be present, then we could ignore the \(b,\bar{b}\) path integration and \(\det(K)\) would be the only contribution to the Casimir force.
This contribution arises from the variable size of the chiral medium, and in more conventional setups with dielectrics this type of Casimir effect is said to arise from ``matching conditions'' between the different media \cite{Bordag:2001qi}.
The impact of \(\det(K)\) on the Casimir force is worked out in \ref{sec:casimir-A}.

To properly formulate the effective boundary theory the action \(S_{b\bar{b}}\) must be rewritten in terms of the \(b,\bar{b}\) fields.
The integrals over \(z\) and \(z^\prime\) can then be performed as the \(b,\bar{b}\) fields do not depend on the \(z,z^\prime\) coordinates.
Moreover \(B_\mu(z)\) and \(\bar{B}_\mu(z)\) contain only terms multiplied by the delta functions \(\delta(z\pm\frac{L}{2})\) and their derivatives
such that integration over \(z\) and \(z^\prime\) merely results in evaluation at \(z,z^\prime=\pm\frac{L}{2}\).
Expanding \(S_{b\bar{b}}\) from \eqref{eq:shifted-action} we get four similar terms
\begin{equation}
    \begin{aligned}
        S_{b\bar{b}} = -\frac{1}{4} \int \frac{\dd[3]{\vec{k}}}{\qty(2\pi)^3} \int \dd{z} \int \dd{z^\prime} \Big[ &B_{\mu}(\vec{k},z) K^{-1}_{\mu\nu}(\vec{k},z,z^\prime) B_\nu(-\vec{k},z^\prime) \\
        + & B_{\mu}(\vec{k},z) K^{-1}_{\mu\nu}(\vec{k},z,z^\prime) \bar{B}_\nu(-\vec{k},z^\prime) \\
        + & \bar{B}_\mu(\vec{k},z) K^{-1}_{\mu\nu}(\vec{k},z,z^\prime) \bar{B}_\nu(-\vec{k},z^\prime) \\
        + & \bar{B}_\mu(\vec{k},z) K^{-1}_{\mu\nu}(\vec{k},z,z^\prime) \bar{B}_\nu(-\vec{k},z^\prime) \Big] \;,
    \end{aligned}
\end{equation}
and concentrating on the \(B_\mu K_{\mu\nu}^{-1} B_\nu\) term we find
\begin{equation}
    \begin{aligned}
        \int \frac{\dd[3]{\vec{k}}}{\qty(2\pi)^3} &\int \dd{z} \int \dd{z^\prime} B_\mu(\vec{k},z) K_{\mu\nu}^{-1}(\vec{k},z,z^\prime) B_{\nu}(-\vec{k},z^\prime) \\
        =\int \frac{\dd[3]{\vec{k}}}{\qty(2\pi)^3} &\Big[ b_i(\vec{k})V_{i\mu}(\vec{k},z) K_{\mu\nu}^{-1}(\vec{k},z,z^\prime) V_{j\nu}(-\vec{k},z^\prime) b_j(-\vec{k}) \\
        & + b_i(\vec{k}) \partial_z \partial_{z^\prime} K_{ij}^{-1}(\vec{k},z,z^\prime) b_j(-\vec{k}) \\
        & - b_i(\vec{k}) \partial_z K_{i\nu}^{-1}(\vec{k},z,z^\prime) V_{j\nu}(-\vec{k},z^\prime) b_j(-\vec{k}) \\
        & - b_i(\vec{k}) V_{i\mu}(\vec{k},z) \partial_{z^\prime} K_{\mu j}^{-1}(\vec{k},z,z^\prime) b_j(-\vec{k}) \eval{\Big]}_{z=z^\prime=\frac{L}{2}} \\
        =\int \frac{\dd[3]{\vec{k}}}{\qty(2\pi)^3} &b_i(\vec{k}) \mathcal{K}_{ij}\qty(\tfrac{L}{2},\tfrac{L}{2}) b_j^\star(\vec{k}) \;,
    \end{aligned}
\end{equation}
where we applied \(b_i(-\vec{k})=b_i^\star(\vec{k})\) and use it to drop the explicit \(\vec{k}\) dependence in the notation where possible, i.e.\ \(b_i(\vec{k})\to b_i\) and \(b_i(-\vec{k})\to b_i^\star\).
From the above expression it follows that \(\mathcal{K}_{ij}\) is defined as
\begin{equation}\label{eq:mathcal-k}
    \begin{aligned}
        \mathcal{K}_{ij}(z,z^\prime) =& V_{i\mu}(\vec{k},z) K_{\mu\nu}^{-1}(\vec{k},z,z^\prime) V_{j\nu}(-\vec{k},z^\prime) + \partial_z \partial_{z^\prime} K_{ij}^{-1}(\vec{k},z,z^\prime) \\
        & - \partial_z K_{i\nu}^{-1}(\vec{k},z,z^\prime) V_{j\nu}(-\vec{k},z^\prime) - V_{i\mu}(\vec{k}, z) \partial_{z^\prime} K_{\mu j}^{-1}(\vec{k},z,z^\prime) \;.
    \end{aligned}
\end{equation}
Repeating this calculation for the \(\bar{B}_\mu K_{\mu\nu}^{-1} B_\nu\), \(B_\mu K_{\mu\nu}^{-1} \bar{B}_\nu\) and \(\bar{B}_\mu K_{\mu\nu}^{-1} \bar{B}_\nu\) terms we find
\begin{equation}
    \begin{aligned}
        \int \frac{\dd[3]{\vec{k}}}{\qty(2\pi)^3} &\int \dd{z} \int \dd{z^\prime} B_\mu(\vec{k},z) K_{\mu\nu}^{-1}(\vec{k},z,z^\prime) \bar{B}_{\nu}(-\vec{k},z^\prime) &= \int \frac{\dd[3]{\vec{k}}}{\qty(2\pi)^3} &b_i \mathcal{K}_{ij}\qty(\tfrac{L}{2},-\tfrac{L}{2}) \bar{b}^\star_j \;, \\
        \int \frac{\dd[3]{\vec{k}}}{\qty(2\pi)^3} &\int \dd{z} \int \dd{z^\prime} \bar{B}_\mu(\vec{k},z) K_{\mu\nu}^{-1}(\vec{k},z,z^\prime) B_{\nu}(-\vec{k},z^\prime) &= \int \frac{\dd[3]{\vec{k}}}{\qty(2\pi)^3} &\bar{b}_i \mathcal{K}_{ij}\qty(-\tfrac{L}{2}, \tfrac{L}{2}) b_j^\star \;, \\
        \int \frac{\dd[3]{\vec{k}}}{\qty(2\pi)^3} &\int \dd{z} \int \dd{z^\prime} \bar{B}_\mu(\vec{k},z) K_{\mu\nu}^{-1}(\vec{k},z,z^\prime) \bar{B}_{\nu}(-\vec{k},z^\prime) &= \int \frac{\dd[3]{\vec{k}}}{\qty(2\pi)^3} &\bar{b}_i \mathcal{K}_{ij}\qty(-\tfrac{L}{2},-\tfrac{L}{2}) \bar{b}_j^\star \;, \\
    \end{aligned}
\end{equation}
i.e.\ only the evaluation of \(\mathcal{K}_{ij}\) at \(z,z^\prime \in \{-\frac{L}{2}, \frac{L}{2}\}\) changes.
The 3D effective theory can consequently be concisely written down as
\begin{equation}
    S_{b\bar{b}}=-\frac{1}{4} \int \frac{\dd[3]{\vec{k}}}{\qty(2\pi)^3}  \mathcal{B}^\dag_i \mathbb{K}_{ij} \mathcal{B}_j \;,
\end{equation}
with the (non-local) matrix kinetic operator \(\mathbb{K}_{ij}\) and the collective \(b,\bar{b}\) fields \(\mathcal{B}\)
\begin{equation}
    \mathbb{K}_{ij} = \mqty(\mathcal{K}_{ij}\qty(\tfrac{L}{2},\tfrac{L}{2}) & \mathcal{K}_{ij}\qty(\tfrac{L}{2},-\tfrac{L}{2}) \\ \mathcal{K}_{ij}\qty(-\tfrac{L}{2},\tfrac{L}{2}) & \mathcal{K}_{ij}\qty(-\tfrac{L}{2},-\tfrac{L}{2})) \qc \mathcal{B}_i=\mqty(b_i(\vec{k}) \;.\\ \bar{b}_i(\vec{k})) \;.
\end{equation}
To bring the action in this form we used the property that \(\mathcal{K}_{ij}(\vec{k},z,z^\prime)=\mathcal{K}_{ji}(-\vec{k},z^\prime,z)\), which follows from the properties of the propagator \(K_{\mu\nu}^{-1}(\vec{k},z,z^\prime)\) and \eqref{eq:mathcal-k}.
The explicit form of \(\mathcal{K}_{ij}\) is best calculated in the \(\tilde{E}\) basis, as there the propagator is a diagonal matrix
\begin{equation}
    K_{rs}^{-1} = \diag(f\qc D^\dag\qc D\qc f) \;,
\end{equation}
and the \(V_{i\mu}(\vec{k},z)\) matrices become
\begin{equation}
    \begin{aligned}
        V_{rs}(z) = \tilde{E}^r_i(\vec{k}) V_{i\mu}(\vec{k},z) \tilde{E}^s_\mu(\vec{k})^\dag &= \mqty(\mqty{\dmat[0]{0 ,-ig\theta(z)\abs{\vec{k}}, ig\theta(z)\abs{\vec{k}}}} & \mqty{i\abs{\vec{k}}\\ 0\\ 0}) \;, \\
        \bar{V}_{rs}(z) = \tilde{E}^r_\mu(\vec{k}) V_{i\mu}(-\vec{k},z) \tilde{E}^s_i(\vec{k})^\dag &= \mqty(\mqty{\dmat[0]{0, -ig\theta(z)\abs{\vec{k}},ig\theta(z)\abs{\vec{k}}} \\ -i\abs{\vec{k}} & 0 & 0}) \;,
    \end{aligned}
\end{equation}
such that \(\mathcal{K}\) becomes
\begin{equation}
    \begin{aligned}
        \mathcal{K}_{rs}(z,z^\prime) =& V_{rq}(z) K_{qt}^{-1}(z,z^\prime) \bar{V}_{ts}(z^\prime) + \partial_z \partial_{z^\prime} K_{rs}^{-1}(z,z^\prime) \\
        & - \partial_z K_{rt}^{-1}(z,z^\prime) \bar{V}_{ts}(z^\prime) - V_{rq}(z) \partial_{z^\prime} K_{q s}^{-1}(z,z^\prime) \;.
    \end{aligned}
\end{equation}
On the other hand, the explicit expression for \(\mathcal{K}_{rs}\) follows as
\begin{equation}
    \begin{aligned}
        \mathcal{K}_{rs}
        &= \mqty(\dmat{\vec{k}^2 f, -g^2\theta(z)\theta(z^\prime) \vec{k}^2 D^\dag, -g^2\theta(z)\theta(z^\prime) \vec{k}^2 D}) + \mqty(\dmat{\partial_z\partial_{z^\prime}f, \partial_z\partial_{z^\prime} D^\dag, \partial_z\partial_{z^\prime} D}) \\
        &\quad - \mqty(\dmat{0, -ig\theta(z^\prime)\abs{\vec{k}}\partial_z D^\dag, ig\theta(z^\prime)\abs{\vec{k}}\partial_z D}) - \mqty(\dmat{0, -ig\theta(z)\abs{\vec{k}}\partial_{z^\prime} D^\dag, ig\theta(z)\abs{\vec{k}}\partial_{z^\prime} D}) \\
        &\implies  \left\{ \begin{array}{ll}
            \mathcal{K}_{00} &= \qty(\partial_z \partial_{z^\prime} + \vec{k}^2)f \\
            \mathcal{K}_{11} &= \qty(\partial_z \partial_{z^\prime} + ig\abs{\vec{k}}\qty(\theta(z)\partial_{z^\prime} + \theta(z^\prime) \partial_{z}) - g^2 \vec{k}^2 \theta(z)\theta(z^\prime)) D^\dag \\
            \mathcal{K}_{22} &= \qty(\partial_z \partial_{z^\prime} - ig\abs{\vec{k}}\qty(\theta(z)\partial_{z^\prime} + \theta(z^\prime) \partial_{z}) - g^2 \vec{k}^2 \theta(z)\theta(z^\prime)) D
            \end{array}
            \right.
        \end{aligned} \;.
\end{equation}
Some care is needed when evaluating \(\mathcal{K}_{rs}\) in \(z=z^\prime=\pm\frac{L}{2}\) though, as when taking derivatives with respect to \(z\) and \(z^\prime\) delta-functions can appear from \(\varphi \subset \bar{D}\) and \(f\):
\begin{equation}
    \begin{aligned}
        \qty(\partial_z \partial_{z^\prime} + \abs{\vec{k}}^2) f(z-z^\prime) &= -\delta(z-z^\prime) \;, \\
        \partial_z \partial_{z^\prime} \varphi(z-z^\prime) &= -\delta(z-z^\prime) - k_c^2 \varphi(z-z^\prime) \;,
    \end{aligned}
\end{equation}
and when \(z=z^\prime=\pm\frac{L}{2}\) this results in \(\delta(0)\), which can be considered equal to zero in the distributional sense, which also becomes evident when using dimensional regularization \cite{Collins:1984xc}. For the record, for any $L\neq0$, $\delta(L)=0$ as well.

Another subtlety is that the first derivatives with respect to \(z\) and \(z^\prime\) of \(f\) and \(\bar{D}\) are discontinuous,
but the combination in which they appear is continuous around \(z,z^\prime \in \qty{-\frac{L}{2},\frac{L}{2}}\)
\begin{equation}
    \begin{aligned}
        \eval{(\theta(z)\partial_{z^\prime}+\theta(z^\prime) \partial_z)\bar{D}(z,z^\prime)}_{z=z^\prime=\pm\frac{L}{2}} &= \frac{2i \theta_\pm}{N^+ N^-} \gamma^+ \gamma^- \sinh(k_c L) \;, \\
        \eval{(\theta(z)\partial_{z^\prime}+\theta(z^\prime) \partial_z)\bar{D}(z,z^\prime)}_{z=\mp\frac{L}{2},z^\prime=\pm\frac{L}{2}} &= -\qty(\theta_+ - \theta_-)\frac{2}{N^+N^-}\frac{k_c}{\abs{\vec{k}}} \;,
    \end{aligned}
\end{equation}
where as before \(\theta_+ \equiv \theta\qty(\frac{L}{2})\) and \(\theta_- \equiv \theta\qty(-\frac{L}{2})\).
The remaining finite (i.e.\ excluding the \(\delta(0)\)) terms evaluated on the plates are
\begin{equation}
    \bar{D}\eval{}_{z=z^\prime=\pm\frac{L}{2}} = \frac{1}{\abs{\vec{k}}} \frac{1}{N^+N^-} \qty(\gamma^+ e^{k_c L} - \gamma^- e^{-k_c L})\qc\bar{D}\eval{}_{z=\pm\frac{L}{2},z^\prime=\pm\frac{L}{2}} = -2 \frac{1}{\abs{\vec{k}}^2} \frac{k_c}{N^+N^-} \;,
\end{equation}
\begin{equation}
        \eval{\partial_z \partial_{z^\prime}\bar{D}}_{z=\pm\frac{L}{2},z^\prime=\mp\frac{L}{2}} = 2 \frac{k_c}{N^+N^-} \;, \eval{\partial_z  \partial_{z^\prime}\bar{D}}_{z=z^\prime=\pm\frac{L}{2}} = \frac{k_c}{N^+N^-} \qty(\gamma^+ e^{L k_c} + \gamma^- e^{-L k_c}) \;.
\end{equation}
With these expressions it follows that
\begin{equation}\label{eq:kpp}
    \begin{aligned}
        \mathcal{K}^{\pm\pm} \equiv& \qty(\partial_z \partial_{z^\prime} - ig\abs{\vec{k}}\qty(\theta\partial_{z^\prime} + \theta^\prime \partial_{z}) - g^2 \vec{k}^2 \theta\theta^\prime) \bar{D}(z, z^\prime)\eval{}_{z=z^\prime=\pm\frac{L}{2}} \;, \\
        =& \frac{\abs{\vec{k}}}{N^+N^-} \qty[\frac{k_c}{\abs{\vec{k}}} \qty(\gamma^+ e^{k_c L} + \gamma^- e^{-k_c L}) \mp 2 i g\gamma^+ \gamma^- \sinh(k_c L) \theta_\pm + g^2 \qty(\gamma^+ e^{k_c L} - \gamma^- e^{-k_c L}) \theta_\pm^2] \;,
    \end{aligned}
\end{equation}
and
\begin{equation}\label{eq:kpn}
    \begin{aligned}
        \mathcal{K}^{\pm\mp} \equiv& \qty(\partial_z \partial_{z^\prime} - ig\abs{\vec{k}}\qty(\theta\partial_{z^\prime} + \theta^\prime \partial_{z}) + g^2 \vec{k}^2 \theta\theta^\prime) \bar{D}(z, z^\prime)\eval{}_{z=\pm\frac{L}{2},z^\prime=\mp\frac{L}{2}} \\
        =& 2\frac{k_c}{N^+N^-} \qty(1 + ig\qty(\theta_+ - \theta_-) + g^2\theta_+ \theta_-) \;.
    \end{aligned}
\end{equation}
Writing the \(b\) and \(\bar{b}\) fields in the \(\tilde{E}\) basis as
\begin{equation}
    b_r = (b_\parallel, b_L, b_R) \qc \bar{b}_r = \qty(\bar{b}_\parallel, \bar{b}_L, \bar{b}_R) \;,
\end{equation}
the $3$D effective action becomes finally
\begin{equation}
    \begin{aligned}
        S_{b\bar{b}}=-\frac{1}{4}\int \frac{\dd[3]{\vec{k}}}{\qty(2\pi)^3} \Big[& b_R \mathcal{K}^{++} b_R^\star + \bar{b}_R \mathcal{K}^{--} \bar{b}_R^\star
        + \bar{b}_R \mathcal{K}^{-+} b_R^\star + b_R \mathcal{K}^{+-} \bar{b}_R^\star %
        + (R \leftrightarrow L) \Big] \\
        = -\frac{1}{4}\int \frac{\dd[3]{\vec{k}}}{\qty(2\pi)^3} \Bigg[&\mathcal{B}^\dag_R \mathbb{K}^R \mathcal{B}_R + \mathcal{B}_L^\dag \mathbb{K}^L \mathcal{B}_L %
        \Bigg] \;,
    \end{aligned}
\end{equation}
where we defined the quantities
\begin{equation}
    \mathcal{B}_R = \mqty(b_R(\vec{k}) \\ \bar{b}_R(\vec{k})) \qc \mathcal{B}_L = \mqty(b_L(\vec{k}) \\ \bar{b}_L(\vec{k})) \qc \mathbb{K}^R = \mqty(\mathcal{K}^{++} & \mathcal{K}^{+-} \\ \mathcal{K}^{-+} & \mathcal{K}^{--}) \qc \mathbb{K}^L = \qty(\mathbb{K}^R)^\star \;.
\end{equation}
The \(b_\parallel\) fields have already been integrated out as this results only in an infinite constant independent of \(L\).
It is now possible to integrate over the \(b\) and \(\bar{b}\) fields, and doing so results in
\begin{equation}
    \begin{aligned}
        \int \DD{\mathcal{B}_R}\DD{\mathcal{B}_L}e^{-S_{b\bar{b}}} &= \frac{C}{\sqrt{\det(\mathbb{K}^R)\det(\qty(\mathbb{K}^R)^\star)}} \;,
    \end{aligned}
\end{equation}
with \(C\) an infinite constant independent of \(L\).
Consequently we have that the (unregularized) vacuum energy is given by
\begin{equation}\label{eq:bbenergy}
    \mathcal{E}_{b\bar{b}} = \frac{1}{2} \int \frac{\dd[3]{\vec{k}}}{\qty(2\pi)^3} \log(\abs{\mathbb{K}^R}\abs{\mathbb{K}^R}^\star) \;,
\end{equation}
where \(\abs{\mathbb{K}^R}\) stands for the matrix determinant of \(\mathbb{K}^R\).

\section{ The Casimir force: various cases}\label{sec:force}

Later on, for mathematical convenience, we will concentrate on calculating the Casimir force which  is the physical quantity of interest:
\begin{equation}
    F = -\dv{\mathcal{E}}{L} \;.
\end{equation}
The energy can be regained, barring a (possibly infinite) constant that may only be important in problems of a gravitational nature, by integrating with respect to $L$.

We will calculate the Casimir force in a few cases and compare with literature where possible.
In section \ref{sec:homogeneous} we look at the easier case where the medium/vacuum is translationally invariant, which allows us to use simpler Green's functions,
while in section \ref{sec:inhomogeneous} we discuss the full case where the chiral medium is only present between the plates.

\subsection{Homogeneous medium}\label{sec:homogeneous}

We first look at the case where the medium is homogeneous, i.e.\ remove the \(z\)-dependence of the parameters in the kinetic part of the action.
In practice this means we take \(\beta(z)\) to be constant \(\beta(z)\equiv \beta\).
In section \ref{sec:cas-qed} we set \(\beta=0\) to arrive at the Casimir effect in QED, but with PEMC conditions.
This serves as a consistency check for our method and can be directly compared with \cite{Dudal:2020yah,Rode:2017yqy}.
Next we look at the chiral QED case with \(\beta\neq 0\) in section \ref{sec:cas-chiral}, which has some interesting differences with the conventional QED Casimir effect.

\subsubsection{The Casimir effect in QED}\label{sec:cas-qed}

\begin{figure}
    \centering
    \includegraphics[width=.7\textwidth]{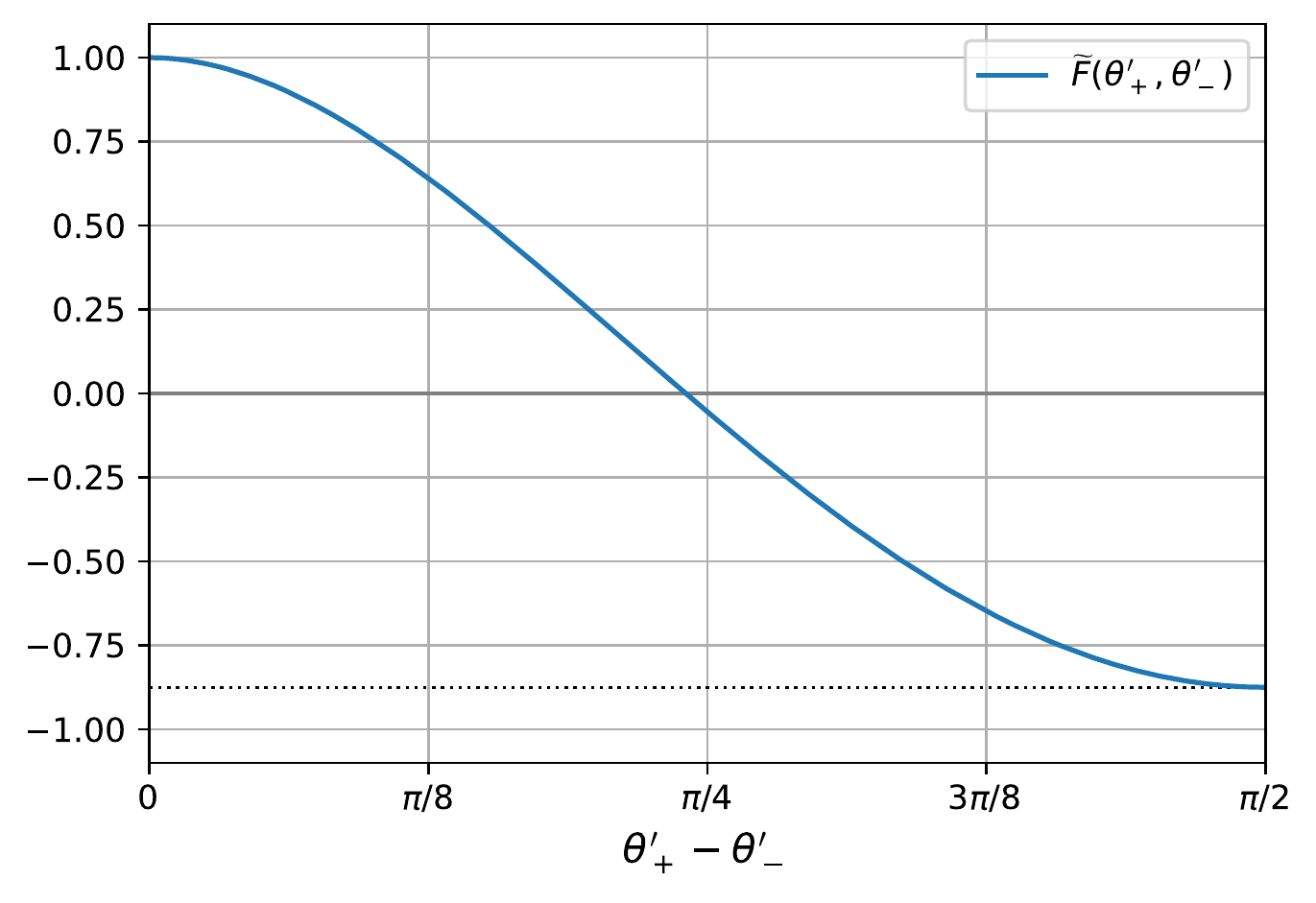}
    \caption{The relative Casimir force \(\tilde{F}_\text{qed}\qty(\theta_+^\prime,\theta_-^\prime)\) in terms of the \textit{duality angles} \(\theta^\prime_\pm = \arctan(g\theta_\pm)\).
    The relative Casimir force only depends on the difference between the duality angles.
    The dotted horizontal line corresponds to \(\tilde{F}=-\frac{7}{8}\).
    The Casimir force vanishes at \(\theta_+^\prime-\theta_-^\prime \approx 0.755\), where it transitions from being attractive to repulsive.}
    \label{fig:qed-casimir}
\end{figure}

Setting \(\beta=0\), the action reduces to that of QED, albeit now with PEMC boundary conditions instead of the usual PEC or PMC conditions.
In this case \eqref{eq:kpp} and \eqref{eq:kpn} become
\begin{equation}
    \begin{aligned}
        \mathcal{K}^{\pm\pm} &= \frac{\abs{\vec{k}}}{2} \qty(1+g^2\theta_\pm^2) \;,\\
        \mathcal{K}^{\pm\mp} &= \frac{\abs{\vec{k}}}{2} \qty(1 + ig\qty(\theta_+ - \theta_-) + g^2 \theta_+ \theta_-) e^{-\abs{\vec{k}}L} \;,
    \end{aligned}
\end{equation}
from which it follows that the determinant of the matrix \(\mathbb{K}^R\) is given by
\begin{equation}
    \abs{\mathbb{K}^R} = \frac{\abs{\vec{k}}^2}{4} \qty[\qty(1+g^2\theta_+^2)\qty(1+g^2\theta_-^2) - \qty(1 + ig\qty(\theta_+ - \theta_-) + g^2 \theta_+ \theta_-)^2 e^{-2\abs{\vec{k}}L}]
\end{equation}
The unregularized Casimir energy is consequently given by
\begin{equation}
    \mathcal{E}_\text{qed} = \frac{1}{2} \int \frac{\dd[3]{\vec{k}}}{\qty(2\pi)^3} \log(\abs{\mathbb{K}^R}\abs{\mathbb{K}^R}^\star) \;.
\end{equation}
The conventional way to regularize the Casimir energy is to subtract the energy at infinite plate separation
\begin{equation}
    \mathcal{E}^\text{reg}_\text{qed} = \frac{1}{2} \int \frac{\dd[3]{\vec{k}}}{\qty(2\pi)^3} \log(\frac{\abs{\mathbb{K}^R}\abs{\mathbb{K}^R}}{\lim\limits_{L\to\infty} \abs{\mathbb{K}^R}\abs{\mathbb{K}^R}})=\frac{1}{2} \int \frac{\dd[3]{\vec{k}}}{\qty(2\pi)^3} \log(\abs{\mathbb{K}^R}_\text{reg}\abs{\mathbb{K}^R}^\star_\text{reg}) \;,
\end{equation}
so that the regularized determinant is given by
\begin{equation}
    \abs{\mathbb{K}^R}_\text{reg} = \frac{\abs{\mathbb{K}^R}}{\lim\limits_{L\to\infty}\abs{\mathbb{K}^R}} = 1 - \frac{1 + ig\qty(\theta_+ - \theta_-) + g^2 \theta_+ \theta_-}{1 - ig\qty(\theta_+ - \theta_-) + g^2 \theta_+ \theta_-}e^{-2\abs{\vec{k}}L} \;.
\end{equation}
The \(\vec{k}\) integral can be done analytically and results in
\begin{equation}
    \mathcal{E}_\text{qed}^\text{reg}(L, \theta_+, \theta_-) = -\frac{1}{8\pi^2 L^3} \Re\qty[\polylog{4}(\frac{1 + ig\qty(\theta_+ - \theta_-) + g^2 \theta_+ \theta_-}{1 - ig\qty(\theta_+ - \theta_-) + g^2 \theta_+ \theta_-})] \;,
\end{equation}
with \(\polylog{n}(x)\) the polylogarithm.
Using that \(\polylog{4}(1) = \frac{\pi^4}{90}\) we can take the PEC and PMC limits
\begin{equation}
    \lim_{\theta_\pm\to\infty}\mathcal{E}_\text{qed}^\text{reg}(L, \theta_+, \theta_-) = \mathcal{E}_\text{qed}^\text{reg}(L, 0, 0) = -\frac{\pi^2}{720 L^3} \equiv \mathcal{E}_\text{qed}(L) \;,
\end{equation}
such that the known equivalence between PEC and PMC boundary conditions for the Casimir energy is recovered.
Similarly we also have that
\begin{equation}
    \mathcal{E}_\text{qed}^\text{reg}(L,\theta,\theta) = \mathcal{E}_\text{qed}(L) \;,
\end{equation}
meaning that the Casimir energy stays the same if the boundary conditions are chosen to be a mixture of PEC and PMC, as long as both plates have the same boundary conditions.
The general Casimir force follows as
\begin{equation}\label{F_qed}
    F_\text{qed}(L,\theta_+,\theta_-) = -\dv{}{L}\mathcal{E}_\text{qed}^\text{reg}(L,\theta_+,\theta_-) = -\frac{3}{8\pi^2 L^4} \Re\qty[\polylog{4}(\frac{1 + ig\qty(\theta_+ - \theta_-) + g^2 \theta_+ \theta_-}{1 - ig\qty(\theta_+ - \theta_-) + g^2 \theta_+ \theta_-})] \;,
\end{equation}
but it is more convenient to instead look at the Casimir force relative to the conventional QED case, which eliminates the \(L\) dependence in this case
\begin{equation}
    \tilde{F}_\text{qed}(\theta_+,\theta_-) = \frac{F_\text{qed}(L,\theta_+,\theta_-)}{F_\text{qed}(L)} \qc F_\text{qed}(L) \equiv F_\text{qed}(L,0,0)= -\frac{\pi^2}{240 L^4} \;.
\end{equation}
Similar to \cite{lindell2005perfect,Rode:2017yqy} we can reparameterize the PEMC boundary conditions as
\begin{equation}
    \cos(\theta_\pm^\prime) F_{\mu 3} + i\sin(\theta_\pm^\prime) \tilde{F}_{\mu 3} \eval_{z=\pm\frac{L}{2}}= 0 \;,
\end{equation}
which can be seen as continuous version of the \(F_{\mu\nu}\leftrightarrow \tilde{F}_{\mu\nu}\) duality transformation applied to the \(F_{\mu 3}\eval_{z=\pm\frac{L}{2}}=0\) boundary conditions.
The \textit{duality angles} \(\theta_\pm^\prime\) relate back to the original parameters as \(\theta_\pm = \tan(g\theta_\pm^\prime)\).
This new parametrization dramatically simplifies the expression for the Casimir force
\begin{equation}
    \begin{aligned}
        \tilde{F}_\text{qed}\qty(\theta_+^\prime, \theta_-^\prime) &= \frac{90}{\pi^4} \Re\polylog{4}(e^{2i\qty(\theta_+^\prime - \theta_-^\prime)}) \\
        &= 1 - \frac{30}{\pi^2} \qty(\theta_+^\prime - \theta_-^\prime)^2 + \frac{60}{\pi^3} \abs{\theta_+^\prime - \theta_-^\prime}^3 - \frac{30}{\pi^4} \qty(\theta_+^\prime - \theta_-^\prime)^4 \;,
    \end{aligned}
\end{equation}
where we have used \cite[eq.~(27.8.6)]{abramowitz1964handbook}, thereby finding the same result as in \cite{Rode:2017yqy,Bordag:1999ux}.
It follows that the Casimir force only depends on the difference \(\theta_+^\prime - \theta_-^\prime\) and is shown in Figure \ref{fig:qed-casimir}.

It can be seen that the maximum amount of repulsion \(-\frac{7}{8}F_\text{qed}(L)\) occurs when \(\theta_+^\prime-\theta_-^\prime=\pm\frac{\pi}{2}\).
In terms of the original parameters this becomes \(\theta_- = -\frac{1}{g^2\theta_+}\).
Writing \(\theta_+ = \theta\) it can be seen that when this is fulfilled the boundary conditions enforced on the plates become
\begin{equation}
    \begin{aligned}
        F_{\mu3} + ig \theta \dual{F}_{\mu 3} \eval{}_{z=-\frac{L}{2}} &= 0 \\
        \dual{F}_{\mu 3} + ig\theta F_{\mu 3} \eval{}_{z=\frac{L}{2}} &= 0 \;,
    \end{aligned}
\end{equation}
which become PEC-PMC (PMC-PEC) boundary conditions in the limit \(\theta\to\infty\) (\(\theta\to 0\)).
It follows that this is a general class of boundary conditions which result in a maximally repulsive (QED) Casimir force, consistent with \cite{Lim:2009zza,PhysRevA.9.2078}.

The \textit{duality angle} formulation allows us to exactly calculate where the Casimir force vanishes
\begin{equation}
    \theta_+^\prime - \theta_-^\prime = \frac{\pi}{2}\qty(1-\sqrt{1-\sqrt{\frac{8}{15}}}) \approx 0.755
\end{equation}
which happens on the transition between the attractive and the repulsive regions of the parameter space.

\subsubsection{The Casimir effect in chiral QED}\label{sec:cas-chiral}

We now move to a chiral medium, and set \(\beta\neq 0\) overall.
As we are still working with a homogeneous medium, we can use the simpler \(\varphi(z-z^\prime)\) propagator instead of \(D(z,z^\prime)\).
Redoing the calculation \eqref{eq:kpp} and \eqref{eq:kpn} results in
\begin{equation}
    \begin{aligned}
        \mathcal{K}^{\pm\pm} &= \frac{k_c}{2} \qty(1 + g^2\frac{\abs{\vec{k}}^2}{k_c^2} \theta_\pm^2) \\
        \mathcal{K}^{\pm\mp} &= \frac{k_c}{2} \qty(1 + ig\frac{\abs{\vec{k}}}{k_c} \qty(\theta_+ - \theta_-) + g^2 \frac{\abs{\vec{k}}^2}{k_c^2}\theta_+ \theta_-) e^{-L k_c} \;,
    \end{aligned}
\end{equation}
from which it follows that the relevant (matrix) determinant is given by
\begin{equation}
    \abs{\mathbb{K}^R} = \frac{1}{4 k_c^2} \qty[\qty(k_c^2 + g^2\abs{\vec{k}}^2\theta_+^2)\qty(k_c^2 + g^2\abs{\vec{k}}^2\theta_-^2) + \qty(k_c + ig\abs{\vec{k}}\theta_+)^2 \qty(k_c - ig\abs{\vec{k}}\theta_-)^2 e^{-2L k_c} ] \;.
\end{equation}
We regularize this determinant analogously to the QED case, giving
\begin{equation}\label{eq:regdet-hom}
    \abs{\mathbb{K}^R}_\text{reg} = 1 - \frac{k_c^2 + ig k_c \abs{\vec{k}} (\theta_+ - \theta_-) + g^2 \abs{\vec{k}}^2 \theta_+ \theta_-}{k_c^2 - ig k_c \abs{\vec{k}} (\theta_+ - \theta_-) + g^2 \abs{\vec{k}}^2 \theta_+ \theta_-} e^{-2 k_c L} \;,
\end{equation}
from which the regularized Casimir energy follows as usual
\begin{equation}
    \mathcal{E}^\text{reg}_\beta\qty(L,\theta_+,\theta_-)= \Re\qty[ \int \frac{\dd[3]{\vec{k}}}{\qty(2\pi)^3} \log(\abs{\mathbb{K}^R}_\text{reg}) ] \;.
\end{equation}
Similarly to the QED Casimir energy, we have that if \(\theta_+=\theta_-\) the \(\theta_\pm\) dependence drops out
\begin{equation}
    \mathcal{E}^\text{reg}_\beta\qty(L,\theta,\theta) \equiv \mathcal{E}_\beta\qty(L)
\end{equation}
for all \(\theta\), which for \(\theta\to 0\) reduces to both plates being PMC and when \(\theta\to\infty\) both plates are PEC.
When both plates enforce the same boundary conditions it follows that the Casimir energy has the form
\begin{equation}\label{eq:chiral-energy}
    \mathcal{E}_\beta\qty(L)= \Re\qty[ \int \frac{\dd[3]{\vec{k}}}{\qty(2\pi)^3} \log(1-e^{-2k_c L}) ] \;,
\end{equation}
and can be calculated analytically (cf.\ Appendix \ref{apx:calculation-chiral-energy})
\begin{equation}
    \mathcal{E}_\beta\qty(L) = \frac{g^4\beta^4 L}{16\pi^2} \sum_{n=1}^\infty \qty[\frac{K_1(g\beta L n)}{g\beta L n}-\frac{K_2(g\beta L n)}{g^2\beta^2 L^2 n^2}] \;,
\end{equation}
where \(K_n(x)\) is the modified Bessel function of the second kind.
The corresponding Casimir force is given by
\begin{equation}
    F_\beta(L) = -\pdv{\mathcal{E}_\beta(L)}{L} = \frac{g^4\beta^4}{16\pi^2} \sum_{n=1}^\infty \qty[K_0(g\beta L n) - 3\frac{K_2(g\beta L n)}{g^2\beta^2 L^2 n^2}] \;,
\end{equation}
which is the same as the Casimir force derived in \cite{Fukushima:2019sjn}.

Our setup has some notable differences with \cite{Fukushima:2019sjn}, however.
A first difference is that we quantize the whole space and consider the plates infinitely thin,
instead of quantizing only the space between plates modelled as semi-infinite slabs.
Our boundary conditions are also gauge invariant, contrary to the boundary condition \(A_\mu=0\) used in \cite{Fukushima:2019sjn}.
Despite these differences we arrive at the same Casimir force.

\begin{figure}
    \centering
    \includegraphics[width=.9\textwidth]{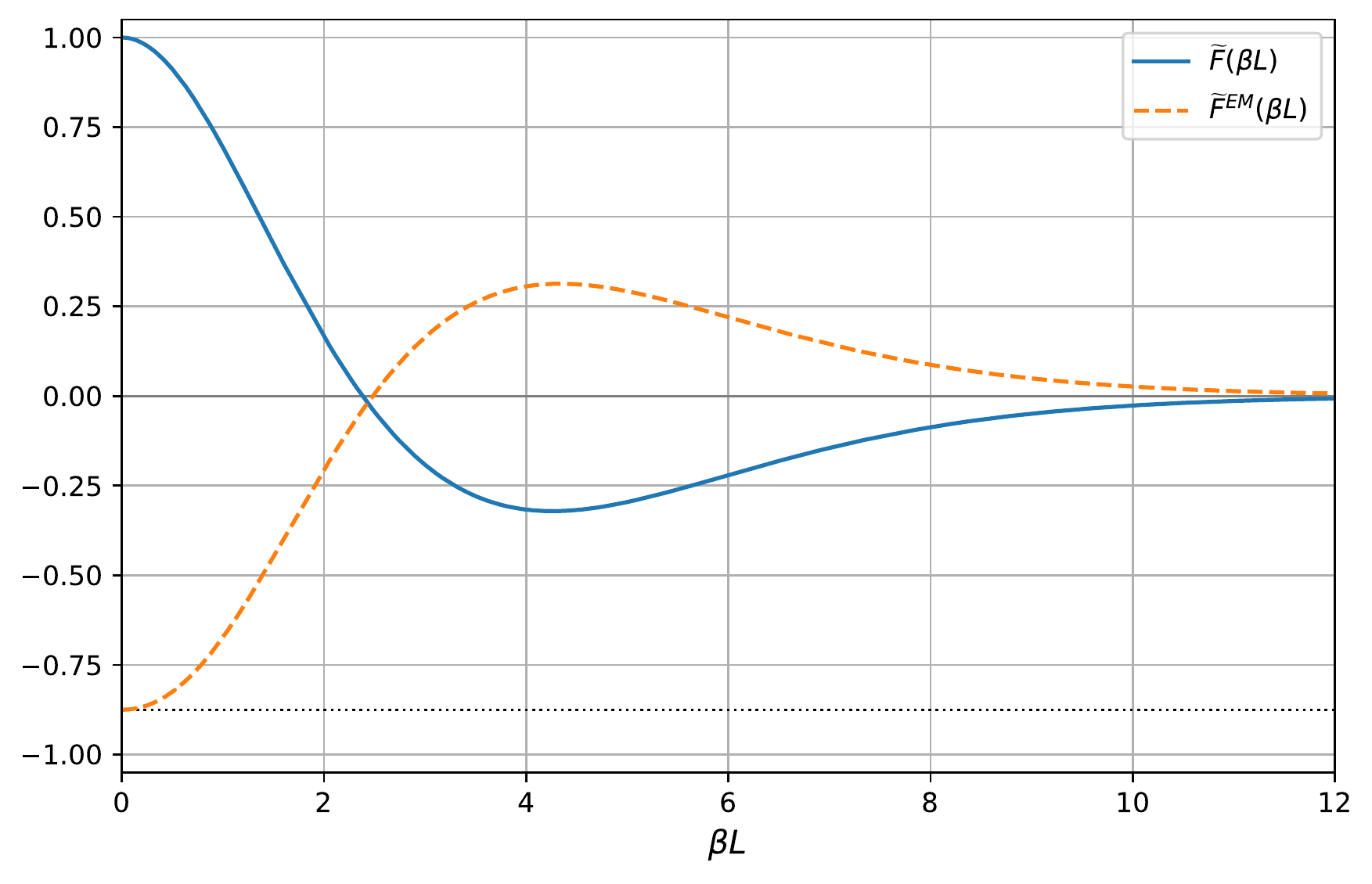}
    \caption{The relative Casimir force \(\tilde{F}(\beta L)\) between two plates with equal boundary conditions in a chiral medium,
    and the case where one plate is PEC and the other is PMC, \(\tilde{F}^{EM}(\beta L)\).
    The dotted line corresponds to \(\tilde{F}=-\frac{7}{8}\).
    When the boundary conditions of the plates are equal the Casimir force starts out equal to the QED Casimir force, but becomes repulsive for larger values of \(\beta L\).
    Similarly in the PEC-PMC case the Casimir force starts repulsive as the QED result of \(-\frac{7}{8}\) times the conventional Casimir force but becomes attractive for larger \(\beta L\).}
    \label{fig:fullchiral}
\end{figure}

Another case where the Casimir energy can be calculated analytically is when one of the plates is PEC while the other is PMC,
which corresponds to the cases \(\theta_\pm\to\infty,\theta_\mp=0\).
The Casimir energy is then given by
\begin{equation}\label{eq:chiral-energy-pec-pmc}
    \begin{aligned}
        \mathcal{E}^{EM}_\beta\qty(L) &= \Re\qty[ \int \frac{\dd[3]{\vec{k}}}{\qty(2\pi)^3} \log(1+e^{-2k_c L}) ] \\
        &= \frac{g^4\beta^4 L}{16\pi^2} \sum_{n=1}^\infty (-1)^n \qty[\frac{K_1(g\beta L n)}{g\beta L n}-\frac{K_2(g\beta L n)}{g^2\beta^2 L^2 n^2}] \;,
    \end{aligned}
\end{equation}
and the Casimir force follows as
\begin{equation}
    F^{EM}_\beta(L) = \frac{g^4\beta^4}{16\pi^2} \sum_{n=1}^\infty (-1)^n \qty[K_0(g\beta L n) - 3\frac{K_2(g\beta L n)}{g^2\beta^2 L^2 n^2}] \;.
\end{equation}
The Casimir force relative to the QED case for the equal boundary conditions \(\tilde{F}\qty(\beta L) = F_\beta(L)/F_\text{qed}(L)\) and the PEC-PMC/PMC-PEC case \(\tilde{F}^{EM}(\beta L) = F^{EM}_\beta(L)/F_\text{qed}(L)\) are shown in Figure \ref{fig:fullchiral}.
It can be seen that the Casimir force in a chiral medium decays quite fast to zero, even with respect to the QED Casimir force, which already decays proportional to \(L^{-4}\).
The chiral Casimir force with equal boundary conditions also displays a repulsive region for \(\beta L\) bigger than \(\sim 2.38\) even though the boundary conditions are invariant under reflection.
In this case the reflection symmetry is broken by the chiral medium itself, namely by the linear gradient of \(\theta(z)\).

Interestingly in the PEC-PMC case the Casimir force starts out as \(\tilde{F}^{EM}(\beta L) \sim -\frac{7}{8} \) for small \(\beta L\), which is the same as the QED result, but becomes attractive for \(\beta L\) larger than \(\sim 2.48\) instead of remaining repulsive.
Consequently \(\beta L \sim 2.48\) is a stable point of the system.

\subsection{Inhomogeneous medium}\label{sec:inhomogeneous}

Moving on to the most interesting (and complicated) case where the chiral medium is present only between the plates,
it follows that there are two contributions to the Casimir force.
One contribution comes from the boundary conditions enforced by the plates,
analogous to the preceding cases.
The other contribution comes from changing the size of the chiral medium,
as the \(\det(K)\) coming from the integration over \(A_\mu\) is now \(L\) dependent. To the best of our knowledge, this case has not yet been worked out in literature.

\subsubsection{The boundary contribution}

In this case we can directly calculate the determinant from \eqref{eq:kpp} and \eqref{eq:kpn}
\begin{equation}
    \abs{\mathbb{K}^R} = \frac{2\abs{k}^2}{N^+N^-} \qty(1 + i g\qty(\theta_+ - \theta_-) + g^2 \theta_+ \theta_-) \qty[\qty(g^2 \theta_+ \theta_- +\frac{k_c^2}{\abs{k}^2})\sinh(L k_c) - i\frac{k_c}{\abs{k}}g \qty(\theta_+ - \theta_-) \cosh(L k_c)] \;,
\end{equation}
which when regularized becomes
\begin{equation}\label{eq:regdet-inhom}
    \abs{\mathbb{K}^R}_\text{reg} = \frac{\abs{\mathbb{K}^R}}{\lim_{L\to\infty}\abs{\mathbb{K}^R}}= \frac{\qty(\gamma^+)^2}{N^+N^-} \qty[e^{k_c L} - \frac{k_c^2 + ig k_c \abs{\vec{k}} (\theta_+ - \theta_-) + g^2 \abs{\vec{k}}^2 \theta_+ \theta_-}{k_c^2 - ig k_c \abs{\vec{k}} (\theta_+ - \theta_-) + g^2 \abs{\vec{k}}^2 \theta_+ \theta_-} e^{-k_c L}] \;,
\end{equation}
where we took into due account the $L$-dependence of $N^{\pm}$.

Similarly to the previous situations the \(\theta_\pm\) dependence drops out when \(\theta_+=\theta_-\),
in which case the regularized determinant becomes
\begin{equation}
    \abs{\mathbb{K}^R}_\text{reg} = \frac{2(\gamma^+)^2}{N^+ N^-}\sinh(L k_c)
\end{equation}
Analogously the PMC-PEC (\(\theta_-=0,\theta_+\to\infty\)) and PEC-PMC (\(\theta_-\to\infty,\theta_+=0\)) cases have
\begin{equation}
    \abs{\mathbb{K}^R}_\text{reg} = \frac{2(\gamma^+)^2}{N^+ N^-}\cosh(L k_c) \;.
\end{equation}
Despite the simpler form of \(\abs{\mathbb{K}^R}_\text{reg}\) when \(\theta_+=\theta_-\) the Casimir energy needs to be calculated numerically
but in the \(\beta\to 0\) limit we arrive at the known QED Casimir force and energy
\begin{equation}
    F_{b\bar{b}}(\beta,L)=-\dv{\mathcal{E}_{b\bar{b}}}{L}\qc F_{b\bar{b}}(\beta=0,L) = -\frac{\pi^2}{240 L^4} \qc \mathcal{E}_{b\bar{b}}^\text{fin}(\beta=0,L) = -\frac{\pi^2}{720 L^3} \;.
\end{equation}

\subsubsection{The \texorpdfstring{\(A_\mu\)}{A} contribution}\label{sec:casimir-A}

The biggest issue with calculating \(\det(K)\), which comes from the \(A_\mu\) integration, is that \(K(\vec{k},z,z^\prime)\) is nondiagonal due to the broken translation symmetry in the \(z\) direction.
Because of this the commonly used identity
\begin{equation}
    \mathcal{E}_A=\frac{1}{2}\log\det(K) = \frac{1}{2}\Tr\log(K) \;,
\end{equation}
is not so useful anymore as the logarithm on the right hand side is no longer a trivial logarithm of the diagonal elements of the operator
\begin{equation}
    \log(K)(\vec{k},z,z^\prime) \neq \log(K(\vec{k},z,z^\prime)) \;,
\end{equation}
and consequently we would need to know the eigenvalues of \(K\) to continue.

Another way of calculating the Casimir energy \(\mathcal{E}_A\) is by treating the coupling to \(\beta(z)\) as an interaction.
The Casimir energy is then given as the sum of all one-loop diagrams.
This calculation has been done in Appendix \ref{apx:oneloop}, and results in a series in \((g\beta)^2\).
The fact that this can be done shows that the Casimir effect is not necessarily proof of vacuum energy,
although the vacuum energy description is the most convenient to use \cite{Jaffe:2005vp}.
A similar construction can be made for boundary conditions on plates, but this requires the boundary conditions to be treated as interactions with a background field \cite{Graham:2002fw}.

It is however much easier to directly calculate the Casimir force instead of the energy
\begin{equation}
    F_A = -\dv{\mathcal{E}_A}{L} \;,
\end{equation}
and use the Jacobi identity to evaluate the derivative with respect to \(L\)
\begin{equation}
    \dv{}{L}\log\det(K) = \det(K)^{-1} \dv{\det(K)}{L} = \Tr[\dv{K}{L} K^{-1}] = \mathcal{V}_2 T \int \dd{z} \int \frac{\dd[3]{\vec{k}}}{\qty(2\pi)^3} \tr[\dv{K}{L} K^{-1}](\vec{k},z,z) \;,
\end{equation}
where \(\mathcal{V}_2 T = \int \dd[3]{\vec{x}}\) is the usual ($3D$) space-time volume coming from the trace.
In the \(\tilde{E}\) basis we have that
\begin{equation}
    K = \diag(\partial_z^2 - \abs{\vec{k}}^2, \partial_z^2 - \qty(k_c^2(z))^\star, \partial_z^2 - k_c^2(z), \partial_z^2-\abs{\vec{k}}^2) \;,
\end{equation}
and the only \(L\) dependence hides inside of the \(k_c^2(z)\) in Equation \eqref{def_k_cz}.
It follows that
\begin{equation}
    \dv{K}{L} = \mqty(\dmat{0, \frac{i}{2}g\beta\abs{\vec{k}}\qty(\delta\qty(z-\frac{L}{2}) + \delta\qty(z+\frac{L}{2})), -\frac{i}{2}g\beta\abs{\vec{k}}\qty(\delta\qty(z-\frac{L}{2}) + \delta\qty(z+\frac{L}{2})), 0}) \;,
\end{equation}
and as such the Casimir force contribution of the \(\det(K)\) term is given by
\begin{equation}
    \begin{aligned}
        F_A &= \frac{1}{2} \int \frac{\dd[3]{\vec{k}}}{\qty(2\pi)^3} \qty[\frac{i}{2}g\beta\abs{\vec{k}}\qty(\bar{D}\qty(\abs{\vec{k}},\frac{L}{2},\frac{L}{2}) + \bar{D}\qty(\abs{\vec{k}},-\frac{L}{2},-\frac{L}{2})) + \qq{c.c.}] \\
        &= -g\beta\int \frac{\dd[3]{\vec{k}}}{\qty(2\pi)^3} \abs{\vec{k}} \Im{\bar{D}\qty(\abs{\vec{k}},\frac{L}{2},\frac{L}{2})} \;,
    \end{aligned}
\end{equation}
where it has been used that \(\bar{D}\qty(\vec{k},\frac{L}{2},\frac{L}{2})=\bar{D}\qty(\vec{k},-\frac{L}{2},-\frac{L}{2})\).
This expression still needs proper regularization, and subtraction of the force at \(L=\infty\) makes this expression finite
\begin{equation}
    F_A^\text{fin} = F_A - F_A\qty(L\to\infty)
\end{equation}
From dimensional analysis it follows that \(F_A\qty(L\to\infty)=g^4\beta^4 F^\infty_A\) with \(F^\infty_A\) a dimensionless constant independent of \(\beta\) and \(L\),
such that this subtraction can be realized by the following counterterm
\begin{equation}
    S_\text{ct} = \int \dd[4]{x} g^4\beta^4(z) F^\infty_A = Lg^4\beta^4 F^\infty_A \int \dd[3]{\vec{x}} = L\beta^4 F^\infty_A \mathcal{V}_2 T \;.
\end{equation}
In other words we have that
\begin{equation}\label{eq:a-regularization}
    \mathcal{E}_A^\text{fin} = \mathcal{E}_A + L F_A(L\to\infty) + C \;,
\end{equation}
with \(C\) an irrelevant (potentially infinite) constant.

\subsubsection{The total Casimir force}

\begin{figure}
    \centering
    \includegraphics[width=.9\textwidth]{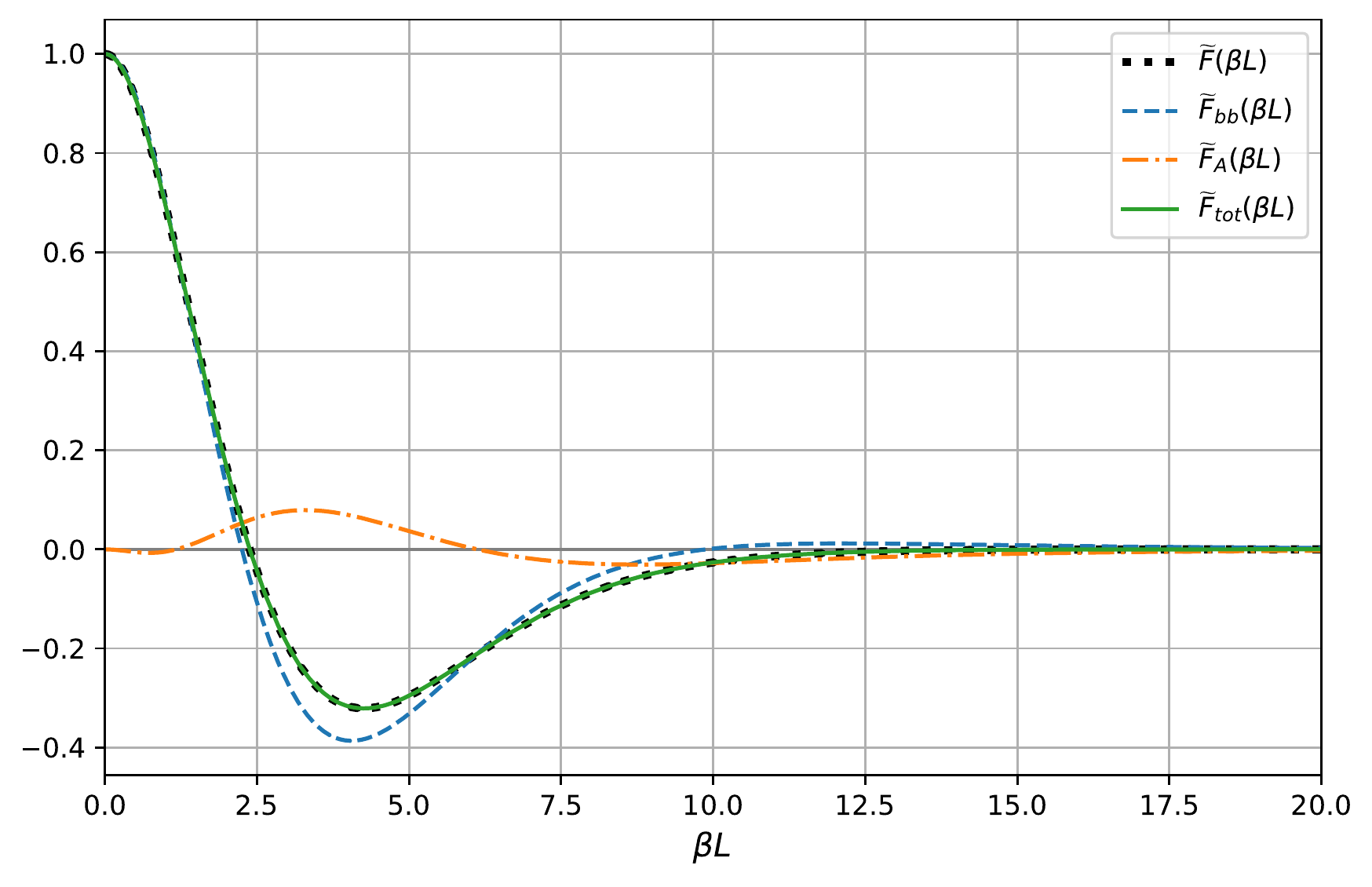}
    \caption{The relative Casimir force \(\tilde{F}_\text{tot}(\beta L)\) and its contributions \(\tilde{F}_A(\beta L)\) and \(\tilde{F}_{b\overline{b}}(\beta L)\) in the case of equal boundary conditions \(\theta_+=\theta_-\).
    The relative Casimir force in a homogeneous chiral medium \(\tilde{F}(\beta L)\) is also shown for reference.}
    \label{fig:allforce}
\end{figure}

\begin{figure}
    \centering
    \includegraphics[width=.9\textwidth]{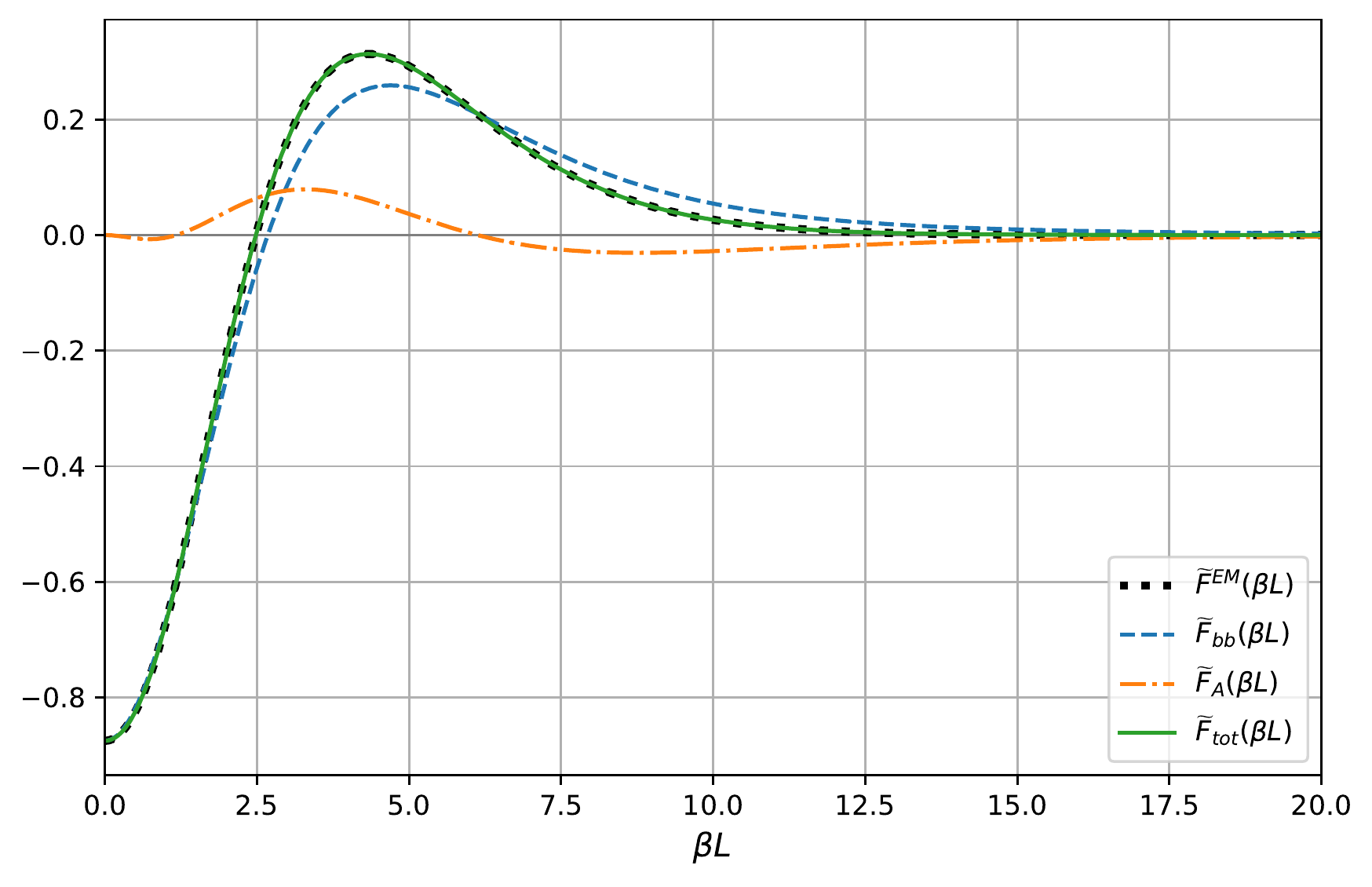}
    \caption{The relative Casimir force \(\tilde{F}_\text{tot}(\beta L)\) and its contributions \(\tilde{F}_A(\beta L)\) and \(\tilde{F}_{b\overline{b}}(\beta L)\) in the case that one plate is PEC and the other is PMC.
    The relative Casimir force in a homogeneous chiral medium \(\tilde{F}^{EM}(\beta L)\) is also shown for reference.}
    \label{fig:pecallforce}
\end{figure}

We can now discuss the total Casimir force relative to the QED Casimir force
\begin{equation}
    \tilde{F}_\text{tot} = \frac{F_\text{tot}}{F_\text{qed}} = \frac{F_{b\bar{b}} + F_A^\text{fin}}{F_\text{qed}} \;.
\end{equation}
From dimensional analysis it follows that the normalized force can only depend on the product \(\beta L\) of the dimensionful parameters.
The total relative Casimir force is shown in Figure \ref{fig:allforce} for the case of equal boundary conditions and in Figure \ref{fig:pecallforce} for PEC-PMC/PMC-PEC boundary conditions,
together with the individual contributions \(\tilde{F}_{b\bar{b}}=F_{b\bar{b}}/F_\text{qed}\) and \(\tilde{F}_A = F_A^\text{fin}/F_\text{qed}\),
and the Casimir force in a chiral medium from the previous section \(\tilde{F}(\beta L)\).
It can be seen that the total Casimir force is numerically the same as the Casimir force in the homogeneous chiral medium,
an equality which also holds for arbitrary values of \(\theta_\pm\).

Comparing the regularized determinants in the homogeneous chiral case \eqref{eq:regdet-hom} and the inhomogeneous case \eqref{eq:regdet-inhom}
it can be seen that they only differ by the factor \(\frac{\qty(\gamma^+)^2}{N^+N^-}e^{k_c L}\).
From the numerical observation that the Casimir force is the same in both situations we can derive an expression for the Casimir energy arising from the inhomogeneous medium
\begin{equation}\label{eq:casimir-energy-a}
    \mathcal{E}_A = \int \frac{\dd[3]{\vec{k}}}{\qty(2\pi)^3} \log(\frac{\qty(\gamma^+)^2}{N^+N^-}e^{k_c L}) \;,
\end{equation}
which was difficult to calculate directly from the functional determinant.
Notice that $\mathcal{E}_A$ is real, as seen from \eqref{F_qed}.
The Casimir force which follows from \(\mathcal{E}_A\) is the same as \(F_A^\text{fin}\),
such that \(\mathcal{E}_A\) is indeed the correctly regularized Casimir energy.

Taking into account that the only constant between section \ref{sec:cas-chiral}, this section, and \cite{Fukushima:2019sjn} is that the medium between the plates is chiral,
it follows that the Casimir force is seemingly insensitive to the medium outside of the plates.
The force \(\tilde{F}_A\) is independent of the boundary conditions however, and persists when the plates are removed,
a feature that is not present in a homogeneous medium.
Moreso \(\tilde{F}_A\) is repulsive at short distances, becomes attractive, and then repulsive again for long distances.
This means that around \(\beta L \approx 1.18\) a chiral medium with variable width is stable, while for \(\beta L \gtrapprox 6.12\) it is unstable and expands until it covers the entire space.

\section{Conclusions and outlook}

From our calculations follows the nontrivial result that the Casimir force is the same whether or not \(\theta(z)\) is constant or remains linear outside of the plates,
while the force itself is consistent with previous calculations \cite{Fukushima:2019sjn}.
The technique developed in \cite{Dudal:2020yah} required some modifications to incorporate a chiral medium which is only present between the plates,
and consequently the nonlocal 3D effective which followed from the boundary conditions did not describe the Casimir effect in its entirety.
It is however remarkable that the homogeneous case results in the same force as the inhomogeneous case, as the former is considerably easier to calculate.

The setup in this paper is however more academic in nature, and can to our knowledge not easily be translated to an experimental setup,
the main obstacle being that either the chiral medium is expected to change in size together with the plates or the plates need to somehow be embedded in a chiral medium.
One possible solution would be for the chiral medium to have a fixed width \(\ell\), independent of the plate separation \(L\).
Such a setup would remove the \(F_A\) contribution, but now \(F_{b\bar{b}}\) would be more complex due to the newly introduced length scale \(\ell\).
This would also provide a connection between the two situations considered in this paper, as \(\ell=L\) corresponds to the chiral medium between the plates (without the \(F_A\) contribution),
and the \(\ell\to \infty\) limit should result in the Casimir effect in the homogeneous chiral medium.

A second work-around would be to ``invert'' our setup from this paper, and have the QED vacuum confined between two semi-infinite slabs of chiral material.
Such a setup looks to be realizable in an experiment using slabs of Weyl semimetals (of sufficient thickness) acting as the plates.
Analogously as in this paper we could manually apply boundary conditions on the boundary between the media.
In this setup the boundary conditions \eqref{eq:organic-boundary-conditions} imply that, as \(\theta=0\) inside of the QED vacuum, the natural boundary conditions would be with \(\theta_\pm = 0\) on their interfaces,
meaning that the interface would act as a perfect magnetic conductor.
Of course we could also once again apply the more general boundary conditions with nonzero \(\theta_\pm\).
The physical case would however be without these manually applied boundary conditions,
and the Casimir force would then arise purely from the matching between the chiral medium and the QED vacuum.
Ultimately the validity of our approach could be checked by comparing our results with \cite{Wilson:2015wsa}.

Another interesting application of our technique would be to study the interface between a topological and a trivial insulator.
Often this interface is chosen to be infinitely sharp and is modeled by letting \(\theta\) jump discontinuously from \(\theta=0\) to \(\theta=\frac{(1+2n)}{4\pi g}\) with \(n\) an integer which depends on non-topological details of the interface \cite{Qi:2008ew}.
Such a jump in \(\theta\) then induces a Chern-Simons term on the interface and is responsible for the Quantum Hall effect.
In practice the transition from a topological to a trivial insulator will occur over a finite distance.
Modelling this transition by letting \(\theta\) be linear would allow us to use the techniques from this paper to determine
what the stable width of such a domain wall could be, assuming the $\theta$ background field is still a valid approximation at these scales.

A step forward could be to use this method in an oscillating $\theta$ background field, simulating axion-like particles (ALPs) coherently coupled with photons in an external (electro-)magnetic field.
This approach is frequently used in the search for ALPs produced in the stars, particularly in the Sun, through the Primakoff effect \cite{Raffelt:1996wa}.
Also, to detect ALPs in the experimental setup known as light shining through the wall in the lab \cite{VanBibber:1987rq}.
Even more, an approach to the one taken in this paper can be used to tackle the problem of stability in solitonic backgrounds in some approximation.

At last, adding finite temperature corrections, see e.g.~\cite{Cruz:2018bqt}, would bring us even closer to realistic predictions of repulsive Casimir forces in chiral media.

\section*{Acknowledgements}
F.~C. has been funded by Fondecyt Grant 1200022. The Centro de Estudios CientÃ­cos (CECs) is funded by the Chilean Government through the Centers of Excellence Base Financing Program of ANID. The work of D.~D.~and T.~O.~was supported by KU Leuven IF project C14/21/087.
P.~P. was funded by Fondo Nacional de Desarrollo Cient\'{i}fico y  Tecnol\'{o}gico--Chile (Fondecyt Grant No.~3200725) and by Charles  University Research Center (UNCE/SCI/013).

\appendix

\section{Green's function coefficients}\label{apx:green-coef}

When \(z^\prime\) lies outside of the chiral medium we have that
\begin{equation}
    \begin{aligned}
        C_1 &= \frac{1}{2 k_c}e^{-k_c \abs{z^\prime}} - \frac{1}{\abs{\vec{k}}N^+} e^{\abs{\vec{k}}\qty(\frac{L}{2} - \abs{z^\prime})} \\
        \eval{C_2}_{z^\prime>\frac{L}{2}} = -\eval{C_2}_{z^\prime<-\frac{L}{2}} &= -\frac{1}{2 k_c} e^{-k_c \abs{z^\prime}} + \frac{1}{\abs{\vec{k}}N^-} e^{\abs{\vec{k}}\qty(\frac{L}{2} - \abs{z^\prime})} \\
        \eval{C^+}_{z^\prime < -\frac{L}{2}} = \eval{C^-}_{z^\prime > \frac{L}{2}} &= \frac{1}{2\abs{\vec{k}}} e^{-\abs{\vec{k}}\abs{z^\prime}} - \frac{2}{N^+ N^-} \frac{k_c}{\abs{\vec{k}}^2} e^{\abs{\vec{k}}\qty(L-\abs{z^\prime})} \\
        \eval{C^-}_{z^\prime < -\frac{L}{2}} = \eval{C^+}_{z^\prime > \frac{L}{2}} &= -\frac{\gamma^+\gamma^-}{N^+N^-} \frac{1}{\abs{\vec{k}}} \sinh(Lk_c) e^{\qty(L-\abs{z^\prime})\abs{\vec{k}}} \;,
    \end{aligned}
\end{equation}
and when \(z^\prime\) lies inside of the medium they take the form
\begin{equation}
    \begin{aligned}
        C_1 &= \frac{\gamma^-}{N^+ k_c} e^{-\frac{L}{2} k_c} \cosh(k_c z^\prime) \;,\\
        C_2 &= \frac{\gamma^-}{N^- k_c} e^{-\frac{L}{2} k_c} \sinh(k_c z^\prime) \;,\\
        C^+ &= \frac{1}{2\abs{\vec{k}}} e^{\abs{\vec{k}} z^\prime} - \frac{1}{\abs{\vec{k}}} e^{\frac{L}{2}\abs{\vec{k}}} \frac{1}{N^+N^-} \qty(\gamma^+ e^{\qty(\frac{L}{2}+z^\prime)k_c} - \gamma^- e^{-\qty(\frac{L}{2}+z^\prime)k_c} ) \;, \\
        C^- &= \frac{1}{2\abs{\vec{k}}} e^{-\abs{\vec{k}} z^\prime} - \frac{1}{\abs{\vec{k}}} e^{\frac{L}{2}\abs{\vec{k}}} \frac{1}{N^+N^-} \qty(\gamma^+ e^{\qty(\frac{L}{2}-z^\prime)k_c} - \gamma^- e^{-\qty(\frac{L}{2}-z^\prime)k_c} ) \;.
    \end{aligned}
\end{equation}

\section{A closed expression for the Casimir energy \texorpdfstring{\(\mathcal{E}_\beta\)}{Eb}}
\label{apx:calculation-chiral-energy}

We start with on the right hand side integral in \eqref{eq:chiral-energy}, therefore
\begin{equation}
    \mathcal{E}_\beta = \Re\qty[\frac{1}{2\pi^{2}} \int_{0}^{+\infty} \dd{\abs{\vec{k}}} \abs{\vec{k}}^{2} \log\qty(1-e^{-2k_c L})] = \frac{1}{4\pi^{2}} (I_{+} + I_{-})\;,
\end{equation}
where we defined
\begin{equation}
    I_{\pm} = \int_{0}^{+\infty} \dd{\abs{\vec{k}}} \abs{\vec{k}}^{2} \log\qty(1-e^{-2\sqrt{\abs{\vec{k}}^{2}\pm ig\beta \abs{\vec{k}}} L}) \;.
\end{equation}
and it was used that \(k_c=\sqrt{\abs{\vec{k}}^2 + ig\beta\abs{\vec{k}}}\).
Writing $\sqrt{\abs{\vec{k}}^2\pm ig\beta \abs{\vec{k}}}=a\pm i b$ and selecting the branch $a>0$, we can see $|e^{-2L\sqrt{\abs{\vec{k}}^2\pm ig\beta \abs{\vec{k}}}}|\leq 1$, allowing us to make the expansion of $\log(1+x)$,
\begin{equation}
    I_ {\pm} = \int_{0}^{+\infty} \dd{\abs{\vec{k}}} \abs{\vec{k}}^2 \sum_{n=1}^{+\infty} \frac{ e^{-2L n \sqrt{\abs{\vec{k}}^2\pm ig\beta \abs{\vec{k}}}} }{n} = \sum_{n=1}^{+\infty} \frac{1}{n} I_{n\pm} \qqtext{where} I_{n\pm} = \int_{0}^{+\infty} \dd{\abs{\vec{k}}} \abs{\vec{k}}^2 e^{-2Ln\sqrt{\abs{\vec{k}}^2\pm ig\beta \abs{\vec{k}}}} \,.
\end{equation}
Let us focus on the integral $I_{n+}$.
The equality $u^{2}+ iu=(u+\frac{i}{2})^{2}+\frac{1}{4}$ suggests us the change of variable $\abs{\vec{k}}=g\beta(z-\frac{i}{2})$, leading to
\begin{equation}
    I_{n+} = g^3 \beta^3 \int_{\frac{i}{2}}^{+\infty+\frac{i}{2}} \dd{z} \qty(z^2 + \frac{i}{2})^2 e^{-2g\beta L n\sqrt{z^2 + \frac{1}{4}}} \;.
\end{equation}
Analogously,
\begin{equation}
    I_{n-} = g^3 \beta^3 \int_{-\frac{i}{2}}^{+\infty-\frac{i}{2}} \dd{z} \qty(z^2 - \frac{i}{2} )^2 e^{-2g\beta L n\sqrt{z^2+\frac{1}{4}}}\;.
\end{equation}
These integrals have paths in the complex plane. Let us relate them to the positive real line path.
To do so, take for $I_{n+}$ the contour $\mathcal{C}_{+}$ when $r\to+\infty$, shown in Figure \ref{fig:complex-plane-fig}.
As there are no poles, and the contour does not cross any branch cut (the branch points are situated at $z=\pm\frac{i}{2}$ and we can take a branch cut lying on the imaginary axis above/below $\pm \frac{i}{2}$.
The contour runs around these branch points along arcs with radius $\epsilon\to 0^+$.
It can be easily checked there is not contribution to the integral.),
because of Cauchy's theorem \cite{ablowitz2003complex}
\begin{equation}
    \int_{\mathcal{C}_+}= \int_{i/2}^{+\infty+ i/2} + I_r + \int_r^0 + I_{1/2} = 0 \;,
\end{equation}
where $I_r$ has the integrand of $I_{n+}$ on the path $z(t)=r - i(t - 1/2)$, $0\leq t \leq 1/2$, namely,
\begin{equation}
    I_{r}=-\int_0^{1/2} i \dd{t} \qty(r - i(t-1) )^2 e^{-2g\beta L n \sqrt{(r+ i(t-1/2))^2 + \frac{1}{4}}} \;,
\end{equation}
and $I_{1/2}$ is the integrand of $I_{n+}$ on the path $z(t)= i t$, $0\leq t \leq 1/2$, i.e.
\begin{equation}
    I_{1/2}=-\int_0^{1/2} \dd{t} \qty(t + \frac{1}{2})^2 e^{-2 g \beta L n \sqrt{-t^2 + \frac{1}{4}}} \;.
\end{equation}
It can be checked that $I_r \to 0$ when $r \to + \infty$. Therefore,
\begin{equation}
    I_{n+}= g^3 \beta^3 \int_0^{+\infty} \dd{x} \qty(x^2 + \frac{i}{2})^2 e^{-2 g \beta L n\sqrt{x^2 + \frac{1}{4}}} - I_{1/2} \;.
\end{equation}
Similarly for $I_{n-}$ but with the contour $\mathcal{C}_-$, when $r \to + \infty$ we obtain
\begin{equation}
    I_{n-}= g^3 \beta^3 \int_0^{+\infty} \dd{x} \qty(x^2 - \frac{i}{2})^2 e^{-2g\beta L n\sqrt{x^2 + \frac{1}{4}}} + I_{1/2} \;.
\end{equation}
Then,
\begin{equation}
    I_{n+}+I_{n-} = 2 g^3 \beta^3 \int_0^{+\infty} \dd{x} x^2 e^{-2g\beta L n\sqrt{x^2 + \frac{1}{4}}} - \frac{1}{2} g^3 \beta^3 \int_0^{+\infty} \dd{x} e^{-2g\beta L n\sqrt{x^2 + \frac{1}{4}}} \;.
\end{equation}
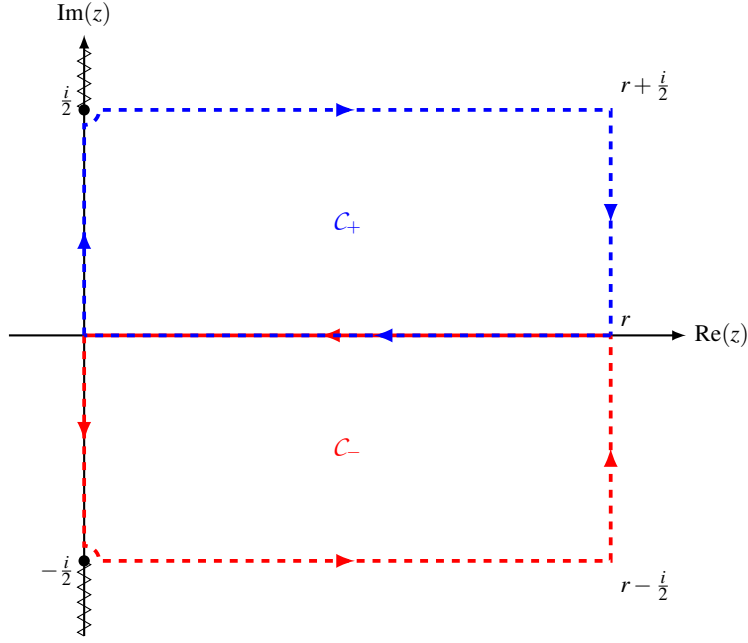
\begin{figure}
    \centering
    \begin{tikzpicture}[
        Cp curve/.style={
            line width= .5 mm,
            color=blue
        },
        Cn curve/.style={
            line width= .5 mm,
            color=red
        },
        arrows/.style={
            postaction={decorate},
            decoration={
                show path construction,
                lineto code={
                    \draw[
                        decorate,
                        decoration={
                            markings,
                            mark=at position #1 with {\arrow{latex}}
                        }
                    ] (\tikzinputsegmentfirst) -- (\tikzinputsegmentlast);
                }
            }
        }
    ]
        \draw[-latex,thick] (-1,0) -- (8,0) node[right] {\(\Re(z)\)};
        \draw[-latex,thick] (0,-4) -- (0,4) node[above] {\(\Im(z)\)};

        \pgfmathsetmacro\eps{.2}
        \draw[Cn curve, dashed, arrows=.5]
            (0,0) -- (0,-3+\eps) node[black, below left] {\(-\frac{i}{2}\)} arc[start angle=90, delta angle=-90, radius=\eps]
                  -- (7,-3) node[black, below right] {\(r - \frac{i}{2}\)}
                  -- (7,0);
        \draw[Cp curve, dashed, arrows=.5]
            (0,0) -- (0,3-\eps) node[black, above left] {\(\frac{i}{2}\)} arc[start angle=-90, delta angle=90, radius=\eps]
                  -- (7,3) node[black, above right] {\(r+\frac{i}{2}\)}
                  -- (7,0) node[black, above right] {\(r\)};

        \draw [Cn curve, arrows=.55, postaction={draw, Cp curve, dashed, arrows=.45}] (7,0) -- (0,0);

        \fill[black] (0,3) circle[radius=.75mm];
        \fill[black] (0,-3) circle[radius=.75mm];

        \draw decorate[decoration={zigzag, segment length=2mm}] {(0,3) -- (0, 3.8)};
        \draw decorate[decoration={zigzag, segment length=2mm}] {(0,-3) -- (0, -4)};

        \node[blue] at (3.5,1.5) {\(\mathcal{C}_+\)};
        \node[red] at (3.5,-1.5) {\(\mathcal{C}_-\)};
    \end{tikzpicture}
    \caption{{\protect\small The two contours in the complex plane.
    Contour $\mathcal{C}_{+}$ goes clockwise passing through $z=0$, $z=\frac{i}{2}$, $z=r+\frac{i}{2}$, and $z=r$.
    Contour $\mathcal{C}_{-}$ goes counterclockwise passing through $z=0$, $z=-\frac{ i}{2}$, $z=r-\frac{i}{2}$, and $z=r$.
    The two branch points at $z=\pm i/2$ are depicted by the black dots. Note that the contours do not cross the branch cuts, which is along the imaginary line.}} %
    \label{fig:complex-plane-fig}
\end{figure}
Once we obtain real integration intervals, these can be written as modified Bessel function of the second kind $K_{\nu}(x)$ (see for instance \cite[p367]{gradshteyn2014table})
\begin{equation}
    \begin{aligned}
        \int\limits_{0}^{+\infty} \dd{x} e^{-a\sqrt{x^2 + b^2 }} &= b K_{1}(ab) \qc (Re(a)>0 \qand Re(b)>0 ) \\
        \int\limits_{0}^{+\infty} \dd{x} x^2 e^{-a\sqrt{x^2 + b^2 }} &= \frac{2b}{a^2} K_1(ab) + \frac{b^2}{a} K_0(ab) = \frac{b^2}{a} K_2(ab) \qc (Re(a)>0 \qand Re(b)>0 )
    \end{aligned}
\end{equation}
where in the last equality we used the identity $K_{2}(x)=\frac{2}{x}\,K_{1}(x) + K_{0}(x)$ (see \cite[p938]{gradshteyn2014table}). In our case $a=g \beta L n$ and $b=\frac{1}{2}$, then
\begin{equation}
    I_+ + I_- = \frac{g^{3}\beta^{3}}{4}\,\left[ \frac{K_{2}(g\beta L n)}{g\beta L n} - K_{1}(g\beta L n) \right] \\
    = \frac{g^{4}\beta^{4}L}{4}\,\sum\limits_{n=1}^{+\infty}\,\left[ \frac{K_{2}(g\beta L n)}{g^{2}\beta^{2} L^{2} n} - \frac{K_{1}(g\beta L n)}{g\beta L } \right]  \;,
\end{equation}
or,
\begin{equation}
    I_+ + I_- = - \frac{g^{4}\beta^{4}L}{4}\,\sum\limits_{n=1}^{+\infty}\,\left[ \frac{K_{2}(g\beta L n)}{g^{2}\beta^{2} L^{2} n^{2}} - \frac{K_{1}(g\beta L n)}{g\beta L n} \right] \;.
\end{equation}
This implies,
\begin{equation}
    \mathcal{E}_\beta = \frac{g^{4}\beta^{4}L}{16\pi^{2}} \, \sum\limits_{n=1}^{+\infty}\,\left[\frac{K_{1}(g\beta L n)}{g\beta L n} - \frac{K_{2}(g\beta L n)}{g^{2}\beta^{} L^{2} n^{2}} \right]\;,
\end{equation}
which is the expression \eqref{eq:chiral-energy}.

In the case that one of the plates is PEC while the other is PMC, the integral expression for the Casimir energy only differs with a sign in the logarithm
\begin{equation}
    \mathcal{E}_\beta^{EM} = \Re\qty[\frac{1}{2\pi^{2}} \int_{0}^{+\infty} \dd{\abs{\vec{k}}} \abs{\vec{k}}^2 \log\qty(1 + e^{-2k_c L})] = \frac{1}{4\pi^{2}} (I_+^{EM} + I_-^{EM})\;.
\end{equation}
Consequently the only difference in the calculation is the appearance of a factor \((-1)^n\) originating from the expansion of the logarithm
\begin{equation}
    I_\pm^{EM} = \sum_{n=1}^\infty \frac{(-1)^n}{n}I_{n\pm}
\end{equation}
The rest of the calculation is the same as in the equal boundary conditions case and results in \eqref{eq:chiral-energy-pec-pmc}.

\section{Casimir energy as sum of one-loop diagrams}
\label{apx:oneloop}

Insead of calculating the Casimir energy resulting from the chiral medium directly we can treat the coupling of the photon to \(\beta(z)\) as an interaction.
It follows that the Casimir energy is then given by the sum of 1PI diagrams.
As the only interaction in our case is with \(\beta(z)\), all the 1PI diagrams (except the pure vacuum diagram) are given by
\begin{equation}
    T\mathcal{V}_2 \mathcal{E}_A =
    \oneloop{photon}{scalar}{1}[.5cm][.5cm][30]
    + \oneloop{photon}{scalar}{2}[.5cm][.5cm][30]
    + \oneloop{photon}{scalar}{3}[.5cm][.5cm][30]
    + \cdots
\end{equation}
where the dashed lines denote the background field \(\beta(z)\) and the photon propagator is simply \(\delta_{\mu\nu}f(\abs{\vec{k}},z-z^\prime)\).
The photon-\(\beta\) vertex corresponds to the insertion of \(g\varepsilon_{ijk}k_k\).
As before the calculations simplify significantly in the \(\tilde{E}\) basis,
and in this basis the diagrams become
\begin{equation}
    \begin{aligned}
        \oneloopn{photon}{scalar}[.5cm][.3cm] = \frac{1}{n}T\mathcal{V}_2\tr[\int \frac{\dd[3]{\vec{k}}}{(2\pi)^3} \int \prod_{i=1}^n \dd{z_i} \mqty(\dmat{-ig\beta(z_i)\abs{\vec{k}}, ig\beta(z_i)\abs{\vec{k}}}) \mqty(\dmat{f(z_i-z_{i+1}),f(z_i-z_{i+1})})] \;,
    \end{aligned}
\end{equation}
where \(z_{n+1}=z_1\) for notational convenience and the factor \(1/n\) is the symmetry factor of the diagram.
Filling in the photon propagator this expression becomes
\begin{equation}
    \oneloopn{photon}{scalar}[.5cm][.3cm] = \frac{2^{1-n}}{n} T\mathcal{V}_2 \Re[i^n] g^n\beta^n (-1)^n \int \frac{\dd[3]{\vec{k}}}{(2\pi)^3} \int_{-\frac{L}{2}}^{\frac{L}{2}}  \qty(\prod_{i=1}^n \dd{z_i}) \exp[-\abs{\vec{k}}\sum_{i=1}^n\abs{z_i - z_{i+1}}] \;,
\end{equation}
where from the presence of \(\Re[i^n]\) it follows that only even \(n\) survive.
Consequently the full expression for the Casimir energy is given by
\begin{equation}
    \mathcal{E}_A = \sum_{n=1}^\infty (-1)^n \frac{1}{n 4^n} \qty(g\beta)^{2n} \int \frac{\dd[3]{\vec{k}}}{(2\pi)^3} \int_{-\frac{L}{2}}^{\frac{L}{2}}  \qty(\prod_{i=1}^{2n} \dd{z_i}) \exp[-\abs{\vec{k}}\sum_{i=1}^{2n}\abs{z_i - z_{i+1}}] \;.
\end{equation}
The \(z_i\) integrals in question are difficult to evaluate for general \(n\), but the first few terms of the Casimir energy are given by
\begin{equation}
    \begin{aligned}
        \mathcal{E}_A =& \int \frac{\dd[3]{\vec{k}}}{(2\pi)^3} \Bigg[ -\frac{(g\beta)^2}{8\abs{\vec{k}}^2}\qty(e^{-2\abs{\vec{k}}L} + 2\abs{\vec{k}}L - 1) \\
        &+ \frac{(g\beta)^4}{256 \abs{\vec{k}}^4} \qty(e^{-4\abs{\vec{k}}L} + 4 e^{-2\abs{\vec{k}}L} \qty(7+10\abs{\vec{k}}L + 4 \abs{\vec{k}}^2 L^2) + 20 \abs{\vec{k}}L - 29) \\
        &- \frac{(g\beta)^6}{6144 \abs{\vec{k}}^6} \Big(e^{-6\abs{\vec{k}}L} + 6e^{-4\abs{\vec{k}}L}\qty(11 + 8 \abs{\vec{k}} L + 8\abs{\vec{k}}^2 L^2) \\
        &\qquad + e^{-2\abs{\vec{k}}L} \qty(495 + 900 \abs{\vec{k}}L + 648 \abs{\vec{k}}^2 L^2 + 224 \abs{\vec{k}}^3 L^3 + 32 \abs{\vec{k}}^4 L^4) + 252\abs{\vec{k}}L - 562 \Big) \\
        &+ \order{(g\beta)^8} \Bigg] \;.
    \end{aligned}
\end{equation}
The \(\vec{k}\) integral is obviously divergent, but the divergent terms are easily recognized as the ones which are not exponentially suppressed for large \(\abs{\vec{k}}\) and \(L\).
This series expansion in \((g\beta)^2\) agrees with the series expansion of \eqref{eq:casimir-energy-a} when the diverging terms are subtracted.
Notice that the diverging terms are at most linear in \(L\), consistent with \eqref{eq:a-regularization}.

\bibliography{bibliography}
\newpage

\end{document}